\documentclass[11pt, prd, aps, showpacs, nofootinbib,superscriptaddress]{revtex4-1}

\usepackage{amsmath,amssymb}
\usepackage{graphicx}
\usepackage{epstopdf}
\usepackage{bm}
\newcommand{\be}{\begin{equation}}
\newcommand{\ee}{\end{equation}}
\newcommand{\bea}{\begin{eqnarray}}
\newcommand{\eea}{\end{eqnarray}}
\allowdisplaybreaks

\newcommand{\Romatre}{Dipartimento di Fisica, Universit\`a  Roma Tre and INFN, Sezione di Roma Tre,\\ Via della Vasca Navale 84, I-00146 Rome, Italy}
\newcommand{\RomatreINFN}{Istituto Nazionale di Fisica Nucleare, Sezione di Roma Tre,\\ Via della Vasca Navale 84, I-00146 Rome, Italy}
\newcommand{\Romadue}{Dipartimento di Fisica and INFN, Universit\`a di Roma ``Tor Vergata",\\ Via della Ricerca Scientifica 1, I-00133 Roma, Italy}
\newcommand{\LaSapienza}{Physics Department and INFN Sezione di Roma La Sapienza,\\ Piazzale Aldo Moro 5, 00185 Roma, Italy}
\newcommand{\soton}{Department of Physics and Astronomy, University of Southampton,\\ Southampton SO17 1BJ, UK}
\newcommand{\Bonn}{HISKP (Theory), Rheinische Friedrich-Wilhelms-Universit\"at Bonn,\\Nussallee 14-16, 53115 Bonn, Germany}

\begin{document}

\title{Comparison of lattice QCD+QED predictions for radiative leptonic decays\\[2mm] of light mesons with experimental data}

\author{R.\,Frezzotti}\affiliation{\Romadue} 
\author{M.\,Garofalo}\affiliation{\Romatre}\affiliation{\Bonn}
\author{V.\,Lubicz}\affiliation{\Romatre} 
\author{G.\,Martinelli}\affiliation{\LaSapienza}
\author{C.T.\,Sachrajda}\affiliation{\soton}
\author{F.\,Sanfilippo}\affiliation{\RomatreINFN}
\author{S.\,Simula}\affiliation{\RomatreINFN}
\author{N.\,Tantalo}\affiliation{\Romadue}

\begin{abstract}
We present a comparison of existing experimental data for the radiative leptonic decays $P\to\ell\nu_\ell\gamma$, where $P=K$ or $\pi$ and $\ell=e$ or $\mu$,
from the KLOE, PIBETA, E787, ISTRA+ and OKA collaborations with theoretical predictions based on the recent non-perturbative determinations of the structure-dependent vector and axial-vector form factors, $F_V$ and $F_A$ respectively. These were obtained using lattice QCD+QED simulations at order $O(\alpha_{\mathrm{em}})$ in the electromagnetic coupling. 
We find good agreement with the KLOE data on $K\to e\nu_e\gamma$ decays from which the form factor $F^+=F_V+F_A$ can be determined. For $K\to\mu\nu_\mu\gamma$ decays
we observe differences of up to 3\,-\,4 standard deviations at large photon energies between the theoretical predictions and the data from the E787, ISTRA+ and OKA experiments and similar discrepancies in some kinematical regions with the PIBETA experiment on radiative pion decays. 
A global study of all the kaon-decay data within the Standard Model results in a poor fit,
largely because at large photon energies the KLOE and E787 data cannot be reproduced simultaneously in terms of the same form factor $F^+$. 
The discrepancy between the theoretical and experimental values of the form factor $F^-=F_V-F_A$ is even more pronounced.
These observations motivate future improvements of both the theoretical and experimental determinations of the structure-dependent form factors $F^+$ and $F^-$, as well as further  
theoretical investigations of models of ``new physics" which might for example, include possible flavor changing interactions beyond $V - A$ and/or non-universal corrections to the lepton couplings.
\end{abstract}

\maketitle

\section{Introduction}
\label{sec:intro}

The decays of charged pseudoscalar mesons into light leptons, $P \to \ell \nu_\ell [\gamma]$ where $\ell$ stands for an electron or a muon, represent an important contribution to flavor physics since they give access to fundamental parameters of the Standard Model (SM), in particular to the Cabibbo-Kobayashi-Maskawa (CKM) matrix elements\,\cite{PDG}.
At tree level, i.e.~without a photon in the final state, these decays are helicity suppressed in the SM due to the $V - A$ structure of the leptonic weak charged current, while the helicity suppression can be overcome by the radiated photons.
Therefore, radiative leptonic decays may provide sensitive probes of possible SM extensions inducing non-standard currents and/or non-universal corrections to the lepton couplings.

Radiative leptonic decays also provide a powerful tool with which to investigate the internal structure of the decaying meson.
In addition to the leptonic decay constant $f_P$, there are other structure-dependent (SD) amplitudes describing the emission of real photons from hadronic states, usually parameterized in terms of the vector and axial-vector form factors, $F_V$ and $F_A$ respectively.
Thus, a first-principle calculation of radiative leptonic decays requires a non-perturbative accuracy, which can be provided by numerical QCD+QED simulations on the lattice.

In Ref.\,\cite{Carrasco:2015xwa} a strategy was proposed to enable lattice computations of QED radiative corrections to $P^+ \to \ell^+ \nu_\ell [\gamma]$ decay rates at order ${\cal{O}}(\alpha_{\mathrm{em}})$.
The strategy naturally obeys the Bloch-Nordsieck mechanism\,\cite{Bloch:1937pw}, in which 
the cancellation of infrared divergences occurs between contributions to the rate with real photons in the final state and those with virtual photons in the decay amplitude. 

Within the RM123 expansion framework\,\cite{deDivitiis:2011eh,deDivitiis:2013xla} the strategy of Ref.\,\cite{Carrasco:2015xwa} was applied in Refs.\,\cite{Giusti:2017dwk,DiCarlo:2019thl} to provide the first non-perturbative model-independent calculation of the SD virtual contribution to the pion and kaon decay rates into muons. 
The contribution with a real photon in the final state was still evaluated in the point-like (pt) effective theory, which is only adequate for sufficiently soft photons (see Ref.\,\cite{Carrasco:2015xwa}).
This limitation has recently been removed in Ref.\,\cite{Desiderio:2020oej}, where the pt and SD amplitudes for real photon emission have been determined non-perturbatively in numerical lattice QCD+QED simulations at order ${\cal{O}}(\alpha_{\mathrm{em}})$ in the electromagnetic coupling. The calculations were performed in the electroquenched approximation in which the sea quarks are electrically neutral\footnote{Note that at order $O(\alpha_{\mathrm{em}})$ the impact of electroquenching on the emission of a real photon is an $SU(3)$-breaking effect, since the contributions from the $u,d$ and $s$ quarks cancel in the SU(3)-symmetric limit.}.

The aim of this work is to carry out a comparison between the theoretical predictions based on the non-perturbative determination of the SD form factors $F_V$ and $F_A$ evaluated in Ref.\,\cite{Desiderio:2020oej} and the experimental data available on the leptonic radiative decay $K \to e \nu_e \gamma$ from the KLOE Collaboration\,\cite{Ambrosino:2009aa}, 
on the decay $K \to \mu \nu_\mu \gamma$ from E787\,\cite{Adler:2000vk}, ISTRA+\,\cite{Duk:2010bs} and OKA\,\cite{Kravtsov:2019amb}~collaborations and 
on the decay $\pi^+ \to e^+ \nu_e \gamma$ from the PIBETA Collaboration\,\cite{Bychkov:2008ws}.

The plan of the remainder of this paper is as follows.
In  Sec.\,\ref{sec:rates} we recall the basic formulae for the double and single differential decay rates $d^{\hspace{1pt}2}\Gamma_1 / dE_\gamma dE_\ell$ and $d\Gamma_1 / d E_\gamma$ for real photon emission, where $E_\gamma$ ($E_\ell$) is the photon (lepton) energy in the rest frame of the decaying meson. The subscript 1 indicates that there is a single photon in the final state.
In  Sec.\,\ref{sec:inclusive} the impact of the SD contributions to the total rates of $\pi_{e2 [\gamma]}$, $\pi_{\mu2 [\gamma]}$, $K_{e2 [\gamma]}$ and $K_{\mu2 [\gamma]}$ decays is evaluated.
We confirm the expectation that the SD contributions to $\Gamma_1$ are negligible for decays into muons, but find that they are a very large contribution to the totally inclusive rate for $K_{e2[\gamma]}$ decays.
In Sections\,\ref{sec:KLOE}-\ref{sec:PIBETA} the experimental data of Refs.\,\cite{Ambrosino:2009aa,Adler:2000vk,Duk:2010bs,Kravtsov:2019amb,Bychkov:2008ws} are briefly described and compared with our theoretical results and with the predictions of Chiral Perturbation Theory (ChPT) at order ${\cal{O}}(e^2 p^4)$\,\cite{Bijnens:1996wm,Geng:2003mt,Mateu:2007tr,Unterdorfer:2008zz,Cirigliano:2011ny}.
For kaon decays, we show that there is a good agreement between our determination of the form factor $F^+$ and the KLOE data on $K\to e\nu_e\gamma$ decays, but we find discrepancies of up to $3$\,-\,$ 4$ standard deviations at large photon energies between our predictions and the  E787, ISTRA+ and OKA data on $K\to\mu\nu_\mu\gamma$ decays. 
We also find similar discrepancies in some kinematical regions of the PIBETA experiment on the radiative pion decay.
In Sec.\,\ref{sec:SM_fit} a simultaneous fit of the KLOE, E787, ISTRA+ and OKA experimental data on radiative kaon decays is performed within the SM and adopting a linear dependence of the SD form factors $F^\pm \equiv F_V \pm F_A$ on the photon energy, as suggested by the lattice results of Ref.\,\cite{Desiderio:2020oej}. 
The quality of the fit is poor because the KLOE and E787 data cannot be reproduced simultaneously in terms of the same form factor $F^+$. 
There is also a particularly pronounced discrepancy between the theoretical and experimental determinations of the form factor $F^-$.
These observation motivate future improved theoretical and experimental determinations of the structure-dependent form factors $F^+$ and $F^-$, as well as further theoretical investigations of theories ``Beyond the Standard Model" which might for example, include possible flavor changing interactions beyond $V - A$ and/or non-universal corrections to the lepton couplings.
Our conclusions are summarized in Sec.\,\ref{sec:conclusions}.

\section{Differential rates for radiative leptonic decays}
\label{sec:rates}

Following Refs.\,\cite{Carrasco:2015xwa,Desiderio:2020oej} the double differential rate for the radiative leptonic decay of a charged pseudoscalar meson, $P^+ \to \ell^+ \nu_\ell \gamma$, can be written as the sum of three contributions:
\be
    \frac{d^2\Gamma(P^+ \to \ell^+ \nu_\ell \gamma)}{dx_\gamma dx_\ell} \equiv  \frac{d^2\Gamma_1}{dx_\gamma dx_\ell} = 
        \frac{\alpha_{\mathrm{em}}}{4\pi} \Gamma^{(0)} ~ \left[ \frac{d^2R_1^{\mathrm{pt}}}{dx_\gamma dx_\ell} +
        \frac{d^2R_1^{\mathrm{SD}}}{dx_\gamma dx_\ell} + \frac{d^2R_1^{\mathrm{INT}}}{dx_\gamma dx_\ell} \right] ~ , ~ 
    \label{eq:dGamma}
\ee
where the subscript $1$ denotes the number of photons in the final state, while $x_\gamma$ and $x_\ell$ are the photon and lepton kinematical variables, defined as
\be
    x_\gamma \equiv \frac{2P \cdot k}{m_P^2} \,,\hspace{0.6in} x_\ell \equiv \frac{2P \cdot p_\ell}{m_P^2} - r_\ell^2\,,
    \label{eq:xgammaxell}
\ee 
where $P$ is the four-momentum of the decaying meson with mass $m_P$, $p_\ell$ is the four-momentum of the final-state lepton with mass $m_\ell$, $k$ is the four-momentum of the photon and $r_\ell \equiv m_\ell / m_P$.
In the rest frame of the decaying meson one has $x_\gamma = 2 E_\gamma / m_P$ and $x_\ell = 2 E_\ell / m_P - r_\ell^2$, where $E_\gamma$ and $E_\ell$ are the photon and lepton energies respectively.

In Eq.\,(\ref{eq:dGamma}) the quantity $\Gamma^{(0)}$ is the leptonic decay rate at tree level, given explicitly by
\be
    \Gamma^{(0)} = \frac{G_F^2 \vert V_{\mathrm{CKM}} \vert^2 f_P^2}{8\pi} m_P^3 r_\ell^2 \left(1 - r_\ell^2 \right)^2 ~ 
    \label{eq:Gamma_QCD}
\ee
where $G_F$ is the Fermi constant, $V_{\mathrm{CKM}}$ the relevant CKM matrix element and $f_P$ the leptonic decay constant of the $P$-meson.

The other entries on the right-hand side of Eq.\,(\ref{eq:dGamma}) are
\bea
    \label{eq:d2Gamma_pt}
    \frac{d^2R_1^{\mathrm{pt}}}{dx_\gamma dx_\ell} & = & \frac{2}{(1 - r_\ell^2)^2} f_{\mathrm{pt}}(x_\gamma, x_\ell)\,, 
    \\[4mm]
    \frac{d^2R_1^{\mathrm{SD}}}{dx_\gamma dx_\ell} & \equiv & \frac{d^2R_1^{\mathrm{SD}^+}}{dx_\gamma dx_\ell} + \frac{d^2R_1^{\mathrm{SD}^-}}{dx_\gamma dx_\ell} \nonumber \\
 &&\hspace{-0.6in}=
        \frac{m_P^2}{2f_P^2 r_\ell^2(1 - r_\ell^2)^2} f_{\mathrm{SD}}^+(x_\gamma, x_\ell) \left[ F^+(x_\gamma) \right]^2  
        +  \frac{m_P^2}{2f_P^2 r_\ell^2(1 - r_\ell^2)^2} f_{\mathrm{SD}}^-(x_\gamma, x_\ell) \left[ F^-(x_\gamma) \right]^2 ,\label{eq:d2Gamma_SD}  \\[5mm]
    \frac{d^2R_1^{\mathrm{INT}}}{dx_\gamma dx_\ell} & \equiv & \frac{d^2R_1^{\mathrm{INT}^+}}{dx_\gamma dx_\ell} + \frac{d^2R_1^{\mathrm{INT}^-}}{dx_\gamma dx_\ell}\nonumber \\
    &&\hspace{-0.6in}= 
        - \frac{2m_P}{f_P\, (1-r_\ell^2)^2} f_{\mathrm{INT}}^+(x_\gamma, x_\ell) F^+(x_\gamma)
         -  \frac{2m_P}{f_P\, (1-r_\ell^2)^2} f_{\mathrm{INT}}^-(x_\gamma, x_\ell) F^-(x_\gamma), 
        \label{eq:d2Gamma_INT}
\eea
where the superscripts $\pm$ correspond to the two photon helicities and the three terms in Eqs.\,(\ref{eq:d2Gamma_pt})-(\ref{eq:d2Gamma_INT}) represent respectively the contribution of the pt approximation of the decaying meson, the SD contribution and the contribution from the interference (INT) between the pt and SD terms. 
Note that in the literature the pt contribution is often referred to as the inner-bremsstrahlung term.
The kinematical functions appearing in Eqs.\,(\ref{eq:d2Gamma_pt})-(\ref{eq:d2Gamma_INT}) are given by
\bea
    \label{eq:fpt}
    f_{\mathrm{pt}}(x_\gamma, x_\ell) & = & \frac{1 - x_\ell}{x_\gamma^2(x_\gamma + x_\ell - 1)} \left[ x_\gamma^2 + 2(1 - x_\gamma)(1 - r_\ell^2) - 
        \frac{2x_\gamma r_\ell^2(1 - r_\ell^2)}{x_\gamma + x_\ell-1} \right] ~ , ~ \\[2mm]
    \label{eq:fSD+}
    f_{\mathrm{SD}}^+(x_\gamma, x_\ell) & = & (x_\gamma + x_\ell - 1) ~ \left[ (x_\gamma + x_\ell - 1 + r_\ell^2)(1 - x_\gamma) - r_\ell^2 \right] ~ , ~ \\[2mm]
    \label{eq:fSD-}
    f_{\mathrm{SD}}^-(x_\gamma, x_\ell) & = & -(1 - x_\ell) ~ \left[ (x_\ell - 1 + r_\ell^2)(1 - x_\gamma) - r_\ell^2 \right] ~ , ~ 
\eea
\bea
    \label{eq:fINT+}
    f_{\mathrm{INT}}^+(x_\gamma, x_\ell) & = & -\frac{1 - x_\ell}{x_\gamma ~ (x_\gamma+x_\ell-1)} \left[ (x_\gamma + x_\ell - 1 + r_\ell^2)(1 - x_\gamma) - 
        r_\ell^2 \right] ~ , ~ \\[2mm]
    \label{eq:fINT-}
    f_{\mathrm{INT}}^-(x_\gamma, x_\ell) & = & \frac{1 - x_\ell}{x_\gamma ~ (x_\gamma + x_\ell - 1)} \left[ x^2_\gamma +
                                                           (x_\gamma + x_\ell - 1 + r_\ell^2)(1 - x_\gamma) - r_\ell^2 \right] ~ , ~
\eea
and the quantities $F^\pm(x_\gamma)$ are the simple combinations 
\be
    \label{eq:Fpm}
    F^\pm(x_\gamma) \equiv F_V(x_\gamma) \pm F_A(x_\gamma) ~ 
\ee
of the vector $F_V(x_\gamma)$ and axial-vector $F_A(x_\gamma)$ form factors which, together with $f_P$, describe the emission of a real photon in the leptonic decay of the $P$-meson. 

Recently the vector and axial-vector form factors have been determined on the lattice for decaying pions, kaons, $D$ and $D_s$ mesons for a wide range of values of $x_\gamma$, adopting the electroquenched approximation in which the sea quarks are electrically neutral\,\cite{Desiderio:2020oej}.
In this work we adopt the definition of the vector ($F_V$) and axial-vector ($F_A$) form factors given in Section II of Ref.\,\cite{Desiderio:2020oej} (see also Appendix B of Ref.\,\cite{Carrasco:2015xwa}). For the decays of the pion and the kaon ($P = \pi, K$) we make use of the linear parameterization of the physical results for $F_V$ and $F_A$ provided in Section V of Ref.\,\cite{Desiderio:2020oej}, 
which is an excellent representation of our lattice data throughout the physical region, i.e.~we write
\bea
    \label{eq:FVA_linear}
    F_V^P(x_\gamma) =  C_V^P + D_V^P \,x_\gamma \,, 
 \qquad
    F_A^P(x_\gamma)  =  C_A^P + D_A^P \, x_\gamma  
\eea
with
\bea
     \label{eq:FV_pion}
     && C_V^\pi = 0.0233 \pm 0.0021 ~ , ~~ D_V^\pi = -0.00026 \pm 0.00027 ~ , ~ \\[2mm] 
     \label{eq:FV_kaon}
     && C_V^K = 0.1244 \pm 0.0096 ~ , ~ \quad D_V^K = -0.024 \pm 0.010 ~ , ~ 
\eea
and
\bea
     \label{eq:FA_pion}
     && C_A^\pi = 0.0104 \pm 0.0026~ , ~~ D_A^\pi = 0.00035 \pm 0.00057 ~ , ~ \\[2mm] 
     \label{eq:FA_kaon}
     && C_A^K = 0.0370 \pm 0.0088 ~ , ~ D_A^K = -0.0012 \pm 0.0074 ~ , ~ 
\eea
where the uncertainties include statistical errors as well as the various sources of systematic errors, except for the QED quenching effect\,\cite{Desiderio:2020oej}.
The impact of the latter is expected to be mild as it is an $SU(3)$-breaking effect.
The full correlation matrices of the parameters in Eqs.\,(\ref{eq:FV_pion})\,-\,(\ref{eq:FA_kaon}) are collected in Tables~\ref{tab:corr_pion} and~\ref{tab:corr_kaon} for pion and the kaon decays respectively.
In the following the uncertainties and correlations of the two form factors are taken into account adopting multivariate gaussian distributions for the parameters in Eqs.\,(\ref{eq:FV_pion})\,-\,(\ref{eq:FA_kaon}) with 10,000 events. 

\begin{table}[!hbt]
\renewcommand{\arraystretch}{1.2}
\begin{center}
\begin{minipage}{0.475\linewidth}	
\begin{tabular}{||c||c|c|c|c||}
\hline
 & ~~~$C_A^\pi$~~~ & ~~~$C_V^\pi$~~~ & ~~~$D_A^\pi$~~~ & ~~~$D_V^\pi$~~~ \\
\hline \hline
 $C_A^\pi$ & ~1.0     & ~0.323 & -0.419  & -0.185 \\ \hline 
 $C_V^\pi$ & ~0.323 & ~1.0     & -0.444  & -0.570 \\ \hline
 $D_A^\pi$ & -0.419  & -0.444  & ~1.0     & ~0.523 \\ \hline
 $D_V^\pi$ & -0.185  & -0.570  & ~0.523 & ~1.0 \\ \hline
\hline
\end{tabular}
\caption{\it \small Correlation matrix for the parameters $C_A^\pi$, $C_V^\pi$, $D_A^\pi$ and $D_V^\pi$ (see Eqs.\,(\ref{eq:FV_pion}) and (\ref{eq:FA_pion})) of the linear parameterization (\ref{eq:FVA_linear}) provided in Ref.\,\cite{Desiderio:2020oej} for the decays of the pion.}
\label{tab:corr_pion}
\end{minipage}
\hfill
\begin{minipage}{0.475\linewidth}	
\begin{tabular}{||c||c|c|c|c||}
\hline
 & ~~~$C_A^K$~~~ & ~~~$C_V^K$~~~ & ~~~$D_A^K$~~~ & ~~~$D_V^K$~~~ \\
\hline \hline
 $C_A^K$ & ~1.0     & ~0.027 & -0.673  & ~0.067 \\ \hline 
 $C_V^K$ & ~0.027 & ~1.0     & ~0.032 & -0.714 \\ \hline
 $D_A^K$ & -0.673  & ~0.032 & ~1.0     & -0.193 \\ \hline
 $D_V^K$ & ~0.067 & -0.714  & -0.193  & ~1.0 \\ \hline
\hline
\end{tabular}
\caption{\it \small Correlation matrix for the parameters $C_A^K$, $C_V^K$, $D_A^K$ and $D_V^K$ (see Eqs.\,(\ref{eq:FV_kaon}) and (\ref{eq:FA_kaon})) of the linear parameterization (\ref{eq:FVA_linear}) provided in Ref.\,\cite{Desiderio:2020oej} for the decays of the kaon.}
\label{tab:corr_kaon}
\end{minipage}
\end{center}
\renewcommand{\arraystretch}{1.0}
\end{table}

The experimental data from the KLOE, E787, ISTRA+, OKA and PIBETA,  collaborations\,\cite{Ambrosino:2009aa,Adler:2000vk,Duk:2010bs,Kravtsov:2019amb,Bychkov:2008ws} correspond to radiative decay rates integrated over the lepton variable $x_\ell$ and including specific kinematical cuts on the lepton momentum and/or on the emission angle $\theta_{\ell \gamma}$ between the lepton and the photon.
We therefore introduce the (partially) {\it integrated} kinematical functions $\widetilde{f}_{\mathrm{pt,SD,INT}}(x_\gamma; x_0, x_1)$ defined as
\bea
       \label{eq:ftildePT}
       \widetilde{f}_{\mathrm{pt}}(x_\gamma; x_0, x_1) & = & \int_{x_0}^{x_1} dx_\ell\, f_{\mathrm{pt}}(x_\gamma, x_\ell)  = - \left[ 2 (1 - r_\ell^2) (1 - x_\gamma) + x_\gamma^2 \right] 
                                                                               \frac{x_1 - x_0}{x_\gamma^2}  \nonumber \\[2mm] 
                    &+& 2 r_\ell^2 (1 - r_\ell^2) \left( \frac{1}{x_\gamma + x_1 - 1} - \frac{1}{x_\gamma + x_0 - 1} \right) 
                                                                               \nonumber \\[2mm]
                    &+& \frac{1}{x_\gamma} \left[ 2 (1 - r_\ell^2)(1 + r_\ell^2 - x_\gamma) + x_\gamma^2 \right] \mbox{log}\left( 
                                                                               \frac{x_\gamma + x_1 - 1}{x_\gamma + x_0 - 1} \right) ~ , ~ \\[2mm] 
      \label{eq:ftildeSD+}
      \widetilde{f}_{\mathrm{SD}}^+(x_\gamma; x_0, x_1) & = & \int_{x_0}^{x_1} dx_\ell\, f_{\mathrm{SD}}^+(x_\gamma, x_\ell)  = - (x_1 - x_0) \Bigl\{ x_\gamma^3 - 
                                                                                   x_\gamma^2 (3 - r_\ell^2 - x_0 - x_1) ~ \nonumber \\[2mm]
                                                                         & - &  \frac{1}{2} r_\ell^2 x_\gamma (2 - x_0 - x_1) + \frac{x_\gamma}{3} \left[ x_1^2 + x_0 x_1 + x_0^2 + 9 - 6 ( x_0 + x_1) \right] ~ 
                                                                                  \nonumber \\[2mm]
                                                                         & + & x_0 + x_1 - 1 - \frac{x_1^2 + x_0 x_1 + x_0^2}{3} \Bigl\} ~ , ~  \\[2mm]
     \label{eq:ftildeSD-}
      \widetilde{f}_{\mathrm{SD}}^-(x_\gamma; x_0, x_1) & = & \int_{x_0}^{x_1} dx_\ell\, f_{\mathrm{SD}}^-(x_\gamma, x_\ell) = - (x_1 - x_0) \Bigl\{ - \frac{1}{2} r_\ell^2 x_\gamma
                                                                                  (2 - x_0 - x_1) ~ \nonumber \\[2mm]
                                                                        & + & \frac{x_\gamma - 1}{3} \left[ x_1^2 + x_0 x_1 + x_0^2 + 3 (1 - x_0 - x_1) \right] \Bigl\} ~ , ~ \\[2mm]
      \label{eq:ftildeINT+}
      \widetilde{f}_{\mathrm{INT}}^+(x_\gamma; x_0, x_1) & = & \int_{x_0}^{x_1} dx_\ell\, f_{\mathrm{INT}}^+(x_\gamma, x_\ell) = - \frac{1}{2} (x_1 - x_0) 
                                                                                    \Bigl[ \frac{2 - x_0 - x_1}{x_\gamma} ~ \nonumber \\[2mm]
                                                                          & - & 2 (1 - r_\ell^2) + x_0 + x_1 \Bigl] + r_\ell^2 x_\gamma  
                                                                                   \mbox{log}\left( \frac{x_\gamma + x_1 - 1}{x_\gamma + x_0 - 1} \right)\,, \\[2mm]
      \label{eq:ftildeINT-}
      \widetilde{f}_{\mathrm{INT}}^-(x_\gamma; x_0, x_1) & = & \int_{x_0}^{x_1} dx_\ell\, f_{\mathrm{INT}}^-(x_\gamma, x_\ell)\nonumber\\[2mm] 
      &=&  - \widetilde{f}_{\mathrm{INT}}^+(x_\gamma, x_0) 
                                                                         -   x_\gamma \left[ x_1 - x_0 - x_\gamma \mbox{log}\left( \frac{x_\gamma + x_1 - 1}{x_\gamma + x_0 - 1} \right) \right],
\eea
where $x_0$ and $x_1$ depend on the specific experimental conditions (see later Sections~\ref{sec:KLOE}-\ref{sec:PIBETA}).
Thus, the partially integrated radiative decay rate for $x_\ell \in [x_0, x_1]$ is given by
\be
     \left[ \frac{d\Gamma_1}{dx_\gamma} \right]_{[x_0, x_1]} = 
         \frac{\alpha_{\mathrm{em}}}{4\pi}\, \Gamma^{(0)} ~ \left\{ \left[ \frac{dR_1^{\mathrm{pt}}}{d x_\gamma} \right]_{[x_0, x_1]} +
         \left[ \frac{dR_1^{\mathrm{SD}}}{d x_\gamma} \right]_{[x_0, x_1]} +\left[ \frac{dR_1^{\mathrm{INT}}}{d x_\gamma} \right]_{[x_0, x_1]} \right\} \,,
    \label{eq:dGamma1}
\ee
where
\bea
    \label{eq:deltaR1_pt_x0_x1}
    \left[ \frac{dR_1^{\mathrm{pt}}}{d x_\gamma} \right]_{[x_0, x_1]} & = & \frac{2}{(1 - r_\ell^2)^2} \widetilde{f}_{\mathrm{pt}}(x_\gamma; x_0, x_1) \,,\\[2mm]
    \label{eq:deltaR1_SD_x0_x1}
    \left[ \frac{dR_1^{\mathrm{SD}}}{d x_\gamma} \right]_{[x_0, x_1]} & = & \frac{m_P^2}{2f_P^2 r_\ell^2(1 - r_\ell^2)^2}
         \left\{ \widetilde{f}_{\mathrm{SD}}^+(x_\gamma; x_0, x_1)  \left[ F^+(x_\gamma) \right]^2 + \widetilde{f}_{\mathrm{SD}}^-(x_\gamma; x_0, x_1) \left[ F^-(x_\gamma) \right]^2 \right\}\,, 
         \qquad \\[2mm]
    \label{eq:deltaR1_INT_x0_x1}
    \left[ \frac{dR_1^{\mathrm{INT}}}{d x_\gamma} \right]_{[x_0, x_1]} & = & - \frac{2 m_P}{f_P(1 - r_\ell^2)^2} 
         \left[ \widetilde{f}_{\mathrm{INT}}^+(x_\gamma; x_0, x_1) F^+(x_\gamma) + \widetilde{f}_{\mathrm{INT}}^-(x_\gamma; x_0, x_1) F^-(x_\gamma) \right]\,.
\eea

In the absence of kinematical cuts $x_\ell$ varies between $x_0 = 1 - x_\gamma + x_\gamma r_\ell^2 / (1 - x_\gamma)$ and $x_1 = 1$, so that in this case
\bea
      \label{eq:fbarpt}
      \widetilde{f}_{\mathrm{pt}} \to \bar{f}_{\mathrm{pt}}(x_\gamma) & = & - \frac{1}{x_\gamma} \Bigl\{ \left[ \frac{(2 - x_\gamma)^2}{1 - x_\gamma} - 4 r_\ell^2 \right] 
                                      (1 - x_\gamma - r_\ell^2) ~ \nonumber \\[2mm]
                                 & - & \left[ 2 (1 - r_\ell^2) (1 + r_\ell^2 - x_\gamma) + x_\gamma^2 \right] 
                                          \mbox{log}\left( \frac{1 - x_\gamma}{r_\ell^2} \right) \Bigl\} ~ , ~ \\[2mm]
     \label{eq:fbarSD+}
      \widetilde{f}_{\mathrm{SD}}^+ \to \bar{f}_{\mathrm{SD}}(x_\gamma) & = & x_\gamma^3 \frac{(2 + r_\ell^2 - 2 x_\gamma)(1 - x_\gamma - r_\ell^2)^2}{6(1 - x_\gamma)^2} ~ , ~ \\[2mm]
       \label{eq:fbarSD-}
      \widetilde{f}_{\mathrm{SD}}^- \to \bar{f}_{\mathrm{SD}}(x_\gamma) & & ~ , ~ \\[2mm]
      \label{eq:fbarINT+}
      \widetilde{f}_{\mathrm{INT}}^+ \to \bar{f}_{\mathrm{INT}}^+(x_\gamma) & = & \frac{x_\gamma}{2} \left[ \frac{r_\ell^4}{1 - x_\gamma} -1 + x_\gamma + 
                                                                                                         2 r_\ell^2 \mbox{log}\left( \frac{1 - x_\gamma}{r_\ell^2} \right) \right] ~ , ~ \\[2mm]
      \label{eq:fbarINT-}
      \widetilde{f}_{\mathrm{INT}}^- \to \bar{f}_{\mathrm{INT}}^-(x_\gamma) & = & - \bar{f}_{\mathrm{INT}}^+(x_\gamma) + x_\gamma^2 \left[ \frac{r_\ell^2}{1 - x_\gamma} -1 + 
                                                                                                        \mbox{log}\left( \frac{1 - x_\gamma}{r_\ell^2} \right) \right] ~ . ~
\eea

\section{Inclusive decay rates for $\boldsymbol{\pi_{\mu(e)2[\gamma]}}$ and 
$\boldsymbol{K_{\mu(e)2[\gamma]}}$ decays}
\label{sec:inclusive}

For real photon emissions the knowledge of the SD vector and axial form factors, $F_V(x_\gamma)$ and $F_A(x_\gamma)$, and of the meson decay constant $f_P$ is sufficient to compute the partially integrated decay rate (\ref{eq:dGamma1}) for any choice of the range of integration $[x_0, x_1]$ over the lepton variable $x_\ell$. 
In this section we consider inclusive decay rates with no kinematical cuts on $x_\ell$ and after integration over the photon variable $x_\gamma$ in its full kinematical range. 

From Eqs.\,(\ref{eq:fbarpt})\,-\,(\ref{eq:fbarINT-}) it can readily be checked that as $x_\gamma \to 0$ one has $dR_1^{\mathrm{SD}} / dx_\gamma \propto x_\gamma^3$ and $dR_1^{\mathrm{INT}} / dx_\gamma \propto x_\gamma$, while $dR_1^{\mathrm{pt}} / dx_\gamma \propto 1 / x_\gamma$.
Therefore, the inclusive SD and INT contributions are infrared safe, while the pt contribution exhibits a logarithmic, structure-independent infrared divergence.
This divergence cancels the corresponding logarithmic infrared divergence of the virtual photon contribution ($\Gamma_0$) to the inclusive decay rate\,\cite{Bloch:1937pw} 
\be
    \label{eq:Gamma_inclusive}
    \Gamma(\Delta E_\gamma) =  \Gamma_0 +  \Gamma_1(\Delta E_\gamma) =  \Gamma_0 +  \int_0^{2 \Delta E_\gamma / m_P} \,  dx_\gamma ~
                                                     \frac{d\Gamma_1}{dx_\gamma} ~ , ~
\ee
where $\Delta E_\gamma$ is the maximum detected energy of the emitted real photon (in the meson rest-frame).
Thus, in the intermediate steps of the calculation of Eq.\,(\ref{eq:Gamma_inclusive}) it is necessary to introduce an infrared regulator.
To this end, a strategy to work only with quantities that are finite when the infrared regulator is removed, has been developed in Ref.\,\cite{Carrasco:2015xwa} and applied to pion and kaon leptonic decays in Refs.\,\cite{Giusti:2017dwk,DiCarlo:2019thl}.
The inclusive rate $\Gamma(\Delta E_\gamma)$ is reorganized as follows
\bea
      \Gamma(\Delta E_\gamma) & = & \displaystyle \lim_{L \to \infty} \left[ \Gamma_0(L) - \Gamma_0^{{\rm pt}}(L) \right] + 
                                                              \displaystyle \lim_{\mu_\gamma \to 0} \left[ \Gamma_0^{{\rm pt}}(\mu_\gamma) + 
                                                              \Gamma_1^{{\rm pt}}(\Delta E_\gamma, \mu_\gamma) \right] ~ \nonumber \\[2mm] 
                                                   & + & \Gamma_1^{\mathrm{SD}}(\Delta E_\gamma) + \Gamma_1^{\mathrm{INT}}(\Delta E_\gamma) ~ 
      \label{eq:Gamma}
 \eea
with the length of the lattice $L$ and $\mu_\gamma$ (for example, a photon mass) acting as infrared regulators in the first two terms on the right-hand side.
The exchange of a virtual photon depends on the structure of the meson since all momentum modes are included, and $\Gamma_0(L)$ must therefore be computed non-perturbatively. We will now explain that on the right-hand side of Eq.\,(\ref{eq:Gamma}), each of the two terms on the top-line are infrared finite, as are separately the two terms on the second line. 

In the first term on the right-hand side of Eq.\,(\ref{eq:Gamma}) the quantities $\Gamma_0(L)$ and $\Gamma_0^{{\rm pt}}(L)$ can be evaluated on the lattice using the lattice size $L$ as the intermediate infrared regulator. 
Both $\Gamma_0(L)$ and  $\Gamma_0^{{\rm pt}}(L)$ have the same infrared divergences which therefore cancel in the difference. 
In our papers we use the lattice size $L$ as the infrared regulator by working in the QED$_\mathrm{L}$ formulation of QED in a finite volume\,\cite{Hayakawa:2008an}, but any other  consistent formulation of QED in a finite volume could equally well be used.  
The difference $\Gamma_0 - \Gamma_0^{{\rm pt}}$ is independent of the regulator as this is removed\,\cite{Lubicz:2016xro}. 
 $\Gamma_0(L)$ depends on the structure of the decaying meson and is computed non-perturbatively on the lattice\,\cite{DiCarlo:2019thl,Lubicz:2016xro}. 

In the second term on the right-hand side of Eq.\,(\ref{eq:Gamma}) the decaying meson is taken to be a point-like charged particle and both $\Gamma_0^{{\rm pt}}(\mu_\gamma)$ and $\Gamma_1^{{\rm pt}}(\Delta E_\gamma, \mu_\gamma)$ are computed  directly in infinite volume, in perturbation theory, using some infrared regulator, for example a photon mass $\mu_\gamma$. 
Each of the two terms is infrared divergent, but the sum is convergent and independent of the regulator\,\cite{Bloch:1937pw}. 
In Refs.\,\cite{Carrasco:2015xwa} and \cite{Lubicz:2016xro} the perturbative calculations of $\left[ \Gamma^{{\rm pt}}_0 + \Gamma^{{\rm pt}}_1(\Delta E_\gamma) \right]$ (see Eq.\,(\ref{eq:IB}) below) and $\Gamma_0^{{\rm pt}}(L)$ have been performed with a small photon mass $\mu_\gamma$ or using the finite volume respectively, as the infrared regulators. 

Each of the two terms on the second line of Eq.\,(\ref{eq:Gamma}) are infrared finite and can be computed directly in infinite volume limit requiring only the knowledge of the
structure dependent form factors,  $F_A(x_\gamma)$ and $F_V(x_\gamma)$, and of the meson decay constant $f_P$\,\cite{Desiderio:2020oej}.

Using the decomposition (\ref{eq:Gamma}), the infrared-finite inclusive decay rate $\Gamma(\Delta E_\gamma)$ can be written as
\be\label{eq:Gamma_final}
    \Gamma(\Delta E_\gamma) = \Gamma^{(0)} \left[ 1 + \delta R_0 + \delta R_{\mathrm{pt}}(\Delta E_\gamma) + \delta R_1^{\mathrm{SD}}(\Delta E_\gamma) + 
                                                    \delta R_1^{\mathrm{INT}}(\Delta E_\gamma) \right]\,, 
\ee
where
\bea\label{eq:deltaR0}
    \delta R_0 & \equiv & \frac{1}{\Gamma^{(0)}} ~ \lim_{L \to \infty} \left[ \Gamma_0(L) - \Gamma_0^{{\rm pt}}(L) \right]\,,\\[2mm]
    \label{eq:deltaR_pt}
    \delta R_{\mathrm{pt}}(\Delta E_\gamma) & \equiv & \lim_{\mu_\gamma \to 0} \left[ \frac{\Gamma_0^{\mathrm{pt}}(\mu_\gamma) + \Gamma_1^{\mathrm{pt}}(\Delta E_\gamma, \mu_\gamma)} 
                                                                        {\Gamma^{(0)}} \right] - 1 \,, \\[2mm]
    \delta R_1^{\mathrm{SD}}(\Delta E_\gamma) & = & \frac{\alpha_{\mathrm{em}}}{4\pi}  \int_0^{\frac{2 \Delta E_\gamma}{m_P}} 
         dx_\gamma ~ \frac{dR_1^{\mathrm{SD}}}{dx_\gamma}\nonumber\\[2mm] 
&&\hspace{-0.6in}= \frac{\alpha_{\mathrm{em}}}{4\pi} 
         \frac{m_P^2}{2f_P^2 r_\ell^2(1 - r_\ell^2)^2} 
         \int_0^{\frac{2 \Delta E_\gamma}{m_P}} dx_\gamma ~ \bar{f}_{\mathrm{SD}}(x_\gamma) \left\{ \left[ F^+(x_\gamma) \right]^2 + 
                         \left[ F^-(x_\gamma) \right]^2 \right\}\,, \label{eq:deltaR1_SD} \\[2mm]
    \delta R_1^{\mathrm{INT}}(\Delta E_\gamma) & = & \frac{\alpha_{\mathrm{em}}}{4\pi} \int_0^{\frac{2 \Delta E_\gamma}{m_P}} 
         dx_\gamma ~ \frac{dR_1^{\mathrm{INT}}}{dx_\gamma} \nonumber \\[2mm]
&&\hspace{-0.6in} =- \frac{\alpha_{\mathrm{em}}}{4\pi}
         \frac{2 m_P}{f_P(1 - r_\ell^2)^2} 
 \int_0^{\frac{2 \Delta E_\gamma}{m_P}} dx_\gamma ~ \left[  \bar{f}_{\mathrm{INT}}^+(x_\gamma) F^+(x_\gamma) + 
                         \bar{f}_{\mathrm{INT}}^-(x_\gamma) F^-(x_\gamma) \right]\,.\label{eq:deltaR1_INT}
\eea
In Eqs.\,(\ref{eq:Gamma_final})\,-\,(\ref{eq:deltaR0}) $\delta R_0$ represents the SD virtual contribution (including also the universal short-distance electroweak correction $(2\alpha_{\mathrm{em}} / \pi) ~ \mbox{log}(M_Z/M_W) \simeq 5.9 \times 10^{-4}$), while in Eq.\,(\ref{eq:deltaR_pt}) $\delta R_{\mathrm{pt}}(\Delta E_\gamma)$ is the (infrared-safe) sum of the point-like contributions of a virtual and a real photon with energy up to $\Delta E_\gamma$, evaluated  
within the $W$-regularization scheme for the ultraviolet divergences which was calculated in Ref.\,\cite{Carrasco:2015xwa} to be
\bea
    \label{eq:IB}
    \delta R_{\mathrm{pt}}(\Delta E_\gamma) & = & \frac{\alpha_{\mathrm{em}}}{4 \pi} \Bigl\{ - 2 \mbox{log}(r_E^2) \left[ 2 + \frac{1 + r_\ell^2}{1 - r_\ell^2} \mbox{log}(r_\ell^2)\right] + 
                                                                3 \mbox{log}\left( \frac{M_P^2}{M_W^2}\right) - 3 \nonumber \\[2mm]
                                                       & + & \frac{3 - 11 r_\ell^2}{1 - r_\ell^2} \mbox{log}(r_\ell^2) - 4 \frac{1 + r_\ell^2}{1 - r_\ell^2} \mbox{Li}_2(1 - r_\ell^2) + 
                                                                 \frac{3 - 6 r_\ell^2 - 4 r_E (1 - r_\ell^2) + r_E^2}{(1 - r_\ell^2)^2} \mbox{log}(1 - r_E) \quad \nonumber \\[2mm]
                                                       & + & \frac{ r_E (4 - 4 r_\ell^2 - r_E)}{(1 - r_\ell^2)^2} \mbox{log}(r_\ell^2) - 4 \frac{1 + r_\ell^2}{1 - r_\ell^2} \mbox{Li}_2(r_E) +
                                                                 \frac{r_E}{2} \frac{22 - 28 r_\ell^2 - 3 r_E}{(1 - r_\ell^2)^2} \Bigl\} ~ , ~
\eea
where $r_E \equiv 2 \Delta E_\gamma / m_P$ and $\mbox{Li}_2(x) = - \int_0^x du ~ \mbox{log}(1 - u) / u$.

Using the vector and axial form factors given in Eqs.\,(\ref{eq:FVA_linear})\,-\,(\ref{eq:FA_kaon}) we have calculated the (totally inclusive) contributions $\delta R_1^{\mathrm{SD}}(\Delta E_\gamma^{max})$ and $\delta R_1^{\mathrm{INT}}(\Delta E_\gamma^{max})$ for the processes $K(\pi) \to \mu(e) \nu \gamma$, where $\Delta E_\gamma^{max} = m_P (1 - r_\ell^2) / 2$.
Our non-perturbative results are shown in Table\,\ref{tab:inclusive} together with the corresponding contribution $\delta R_{\mathrm{pt}}(\Delta E_\gamma^{max})$ from Eq.\,(\ref{eq:IB}).
For the ratio ($m_P / f_P$) appearing in Eqs.\,(\ref{eq:deltaR1_SD})\,-\,(\ref{eq:deltaR1_INT}) we take the values ($139.6~{\rm MeV} / 130.4~{\rm MeV}$) and ($493.7~{\rm MeV} / 156.1~{\rm MeV}$) for $P = \pi$ and $K$, respectively\,\footnote{For the kaon the value $f_K = 156.1$ MeV is taken from Ref.\,\cite{DiCarlo:2019thl} and is based on the latest FLAG average\,\cite{Aoki:2019cca} for $f_{K^+}$ corrected for strong $SU(2)$ breaking effects.}.

\begin{table}[!hbt]
\renewcommand{\arraystretch}{1.2}	 
\begin{center}	
\begin{tabular}{||c||c|c||c|c||}
\hline
 & $\pi_{e2[\gamma]}$ & $\pi_{\mu2[\gamma]}$ & $K_{e2[\gamma]}$ & $K_{\mu2[\gamma]}$\\
\hline \hline
$\delta R_0$                                                   & $^{(*)}$ & $0.0411~(19)$ & $^{(*)}$ & $0.0341~(10)$\\ \hline 
$\delta R_{\mathrm{pt}}(\Delta E_\gamma^{max})$       & $-0.0651$ & $-0.0258$ & $-0.0695$ & $-0.0317$\\ \hline
$\delta R_1^{\mathrm{SD}}(\Delta E_\gamma^{max})$ & $5.4~(1.0) \times 10^{-4}$ & $2.6~(5) \times 10^{-10}$ & $1.19~(14)$ & $2.2~(3) \times 10^{-5}$\\ \hline
$\delta R_1^{\mathrm{INT}}(\Delta E_\gamma^{max})$ & $-4.1~(1.0) \times 10^{-5}$ & $-1.3~(1.5) \times 10^{-8}$ & $-9.2~(1.3) \times 10^{-4}$ & $-6.1~(1.1) \times 10^{-5}$\\ \hline \hline
$\Delta E_\gamma^{max}$  (MeV)                 & $69.8$ & $29.8$ & $246.8$ & $235.5$\\ 
\hline \hline
\end{tabular}
\end{center}
\renewcommand{\arraystretch}{1.0}
\begin{minipage}{\linewidth}
\begin{itemize}
   \item[$^{(*)}$] {\it Not yet evaluated by numerical lattice QCD+QED simulations.}
\end{itemize}
\end{minipage}
\caption{\it \small Values of the contributions $\delta R_0$, $\delta R_{\mathrm{pt}}(\Delta E_\gamma^{max})$, $\delta R_1^{\mathrm{SD}}(\Delta E_\gamma^{max})$ and $\delta R_1^{\mathrm{INT}}(\Delta E_\gamma^{max})$, defined in Eqs.\,(\ref{eq:deltaR0})-(\ref{eq:deltaR1_INT}), evaluated using the lattice results of Refs.\,\cite{DiCarlo:2019thl,Desiderio:2020oej} for the decays $K(\pi) \to \mu(e) \nu [\gamma]$. In the last row the values of the maximum photon energy, $\Delta E_\gamma^{max}$, are also shown for each decay process.}
\label{tab:inclusive}
\end{table} 

In the same Table we also show the values of the SD virtual contributions $\delta R_0(\pi_{\mu2})$ and $\delta R_0(K_{\mu2})$, which can be derived from the results of Ref.\,\cite{DiCarlo:2019thl}.
There, the combination $\delta R_0 + \delta R_{\mathrm{pt}}(\Delta E_\gamma^{max})$ was evaluated for $K(\pi) \to \mu \nu [\gamma]$ decays, obtaining 
\bea
    \label{eq:deltaR_pi}
     \delta R_0(\pi_{\mu2}) + \delta R_{\mathrm{pt}}(\pi_{\mu2[\gamma]}; \Delta E_\gamma^{max}) & = & 0.0153 ~ (19) ~ , ~ \\[2mm]
     \label{eq:deltaR_K}
     \delta R_0(K_{\mu2}) + \delta R_{\mathrm{pt}}(K_{\mu2[\gamma]}; \Delta E_\gamma^{max}) & = & 0.0024 ~ (10) ~ . ~
\eea
For decays into a final-state electron, the lattice determinations of the SD virtual contributions $\delta R_0(\pi_{e2})$ and $\delta R_0(K_{e2})$, which are currently missing in Table\,\ref{tab:inclusive}, are in progress.

From Table\,\ref{tab:inclusive} it can be seen that for radiative decays into muons the SD and INT contributions are negligible compared to the pt one, and, therefore, the results (\ref{eq:deltaR_pi}) and~(\ref{eq:deltaR_K}) represent respectively the totally inclusive corrections to the tree-level decay of pions and kaons into muons.
This had been anticipated in Ref.\,\cite{DiCarlo:2019thl}, where the SD and INT contributions were neglected in the  
extraction of the CKM matrix element $|V_{us}|$ using the experimental result for the total decay rate $\Gamma(K \to \mu \nu [\gamma])$ from the PDG\,\cite{PDG}.  

The situation is very different for radiative kaon decays into electrons where the relative SD contribution is very large and even exceeds $1$.
This is related to the presence of the factor $r_\ell^2$ in the denominator of Eq.\,(\ref{eq:deltaR1_SD}), which compensates the factor $r_\ell^2$ present in the tree-level rate $\Gamma^{(0)}$ because of helicity suppression (see Eq.\,(\ref{eq:Gamma_QCD})).
In the next Section we will compare our non-perturbative predictions with results from the KLOE experiment on the radiative kaon decay $K_{e2\gamma}$, which is devoted to the investigation of this large SD contribution\,\cite{Ambrosino:2009aa}.

The discussion and results in this section concerned the rates for inclusive decays to which the exchange of a virtual photon contributes significantly. For the remainder of this paper we focus on the differential rates for decays with a real photon in the final state, i.e. $P\to\ell\nu_\ell\gamma$ decays.

\section{Comparison with the experimental results from the KLOE collaboration}
\label{sec:KLOE}

In Ref.\,\cite{Ambrosino:2009aa} the KLOE Collaboration has measured the differential decay rate $d\Gamma(K_{e2\gamma}) / dE_\gamma$ for photon energies in the range $10~{\rm MeV} < E_\gamma < E_\gamma^{max} \simeq 250~{\rm MeV}$ with the constraint $p_e > 200$ MeV.
More precisely, they have measured the differential branching ratio
\be
    \label{eq:width}
    \frac{dR^{\mathrm{exp}}}{d E_\gamma} = \frac{1}{\Gamma(K _{\mu2[\gamma]})} \left[ \frac{d\Gamma(K _{e2\gamma})}{d E_\gamma} \right]_{p_e > 200 {\rm MeV}}
\ee
integrated in five different bins of photon energies:
\be
   \label{eq:DeltaR}
    \Delta R^{\mathrm{exp}, i} \equiv \int_{E_\gamma^i}^{E_\gamma^{i+1}} d E_\gamma ~ \frac{dR^{\mathrm{exp}}}{d E_\gamma}
\ee
with $E_\gamma^i = \{ 10, 50, 100, 150, 200, 250 \}$\,MeV.

Since we work at first order in $\alpha_{\mathrm{em}}$, we can replace $\Gamma(K _{\mu2[\gamma]})$ with its tree-level expression (\ref{eq:Gamma_QCD}) in the denominator of Eq.\,(\ref{eq:width})\,\footnote{The results shown in Table\,\ref{tab:inclusive} imply that the difference between the total rate $\Gamma(K _{\mu2[\gamma]})$ and its tree-level expression $\Gamma^{(0)}(K_{\mu2})$  is at the level of few permille.}.
Thus, the theoretical prediction $\Delta R^{\mathrm{th}, i}$ can be decomposed into the sum of three terms
\be
    \label{eq:DeltaRdec}
    \Delta R^{\mathrm{th}, i} = \Delta R^{\mathrm{pt}, i} + \Delta R^{\mathrm{SD}, i} + \Delta R^{\mathrm{INT}, i} ~ , ~
\ee
where
\be
   \label{eq:DeltaRi}
    \Delta R^{\mathrm{pt\,(SD,INT)}, i} = \frac{\Gamma^{(0)}(K_{e2})}{\Gamma^{(0)}(K_{\mu2})} ~ \int_{2E_\gamma^i/m_K}^{2E_\gamma^{i+1}/m_K} d x_\gamma ~ 
                                             \left[ \frac{dR_1^{\mathrm{pt\,(SD,INT)}}}{d x_\gamma} \right]_{p_e > 200 {\rm MeV}}
\ee
with 
\be
    \frac{\Gamma^{(0)}(K_{e2})}{\Gamma^{(0)}(K_{\mu2})} =  \frac{m_e^2}{m_\mu^2} \frac{(1 - r_e^2)^2}{(1 - r_\mu^2)^2} 
                                                                                                \simeq 2.5689 \times 10^{-5} ~ 
\ee
and $r_e = m_e / m_K$ and $r_\mu = m_\mu / m_K$.

The presence of a constraint of the type $p_e > p_{e, min}$ implies that $x_e > x_{min}$, where $x_{min}$ is given by
\be
     \label{eq:xmin}
     x_{min} = \frac{2}{m_K} \sqrt{m_e^2 + p_{e, min}^2} - r_e^2 ~  . ~
\ee
We therefore obtain  
\bea
    \label{eq:deltaR1_pt_x0}
    \left[ \frac{dR_1^{\mathrm{pt}}}{d x_\gamma} \right]_{p_e > p_{e, min}} & = & \frac{\alpha_{\mathrm{em}}}{4\pi} \frac{2}{(1 - r_e^2)^2} 
        \widetilde{f}_{\mathrm{pt}}(x_\gamma; x_0, 1) ~ , ~ \\[2mm]
    \label{eq:deltaR1_SD_x0}
    \left[ \frac{dR_1^{\mathrm{SD}}}{d x_\gamma} \right]_{p_e > p_{e, min}} & = & \frac{\alpha_{\mathrm{em}}}{4\pi} \frac{m_K^2}{2f_K^2 r_e^2(1 - r_e^2)^2}
         \left\{ \widetilde{f}_{\mathrm{SD}}^+(x_\gamma; x_0, 1)  \left[ F^+(x_\gamma) \right]^2 \right. ~ \nonumber \\[2mm]
         & + & \left. \widetilde{f}_{\mathrm{SD}}^-(x_\gamma; x_0, 1) \left[ F^-(x_\gamma) \right]^2 \right\} ~ , \quad \\[2mm]
    \label{eq:deltaR1_INT_x0}
     \left[ \frac{dR_1^{\mathrm{INT}}}{d x_\gamma} \right]_{p_e > p_{e, min}} & = & - \frac{\alpha_{\mathrm{em}}}{4\pi} \frac{2 m_K}{f_K(1 - r_e^2)^2} 
         \left[ \widetilde{f}_{\mathrm{INT}}^+(x_\gamma; x_0, 1) F^+(x_\gamma) \right. ~ \nonumber \\[2mm]
        & + & \left. \widetilde{f}_{\mathrm{INT}}^-(x_\gamma; x_0, 1) F^-(x_\gamma) \right] ~  
\eea
where $x_0$ is given by
\be
     \label{eq:x0}
     x_0 \equiv {\rm max}\left(x_{min}, 1 - x_\gamma + x_\gamma \frac{r_e^2}{1 - x_\gamma} \right) ~ . ~  
\ee

\begin{figure}[htb!]
\begin{center}
\includegraphics[scale=0.85]{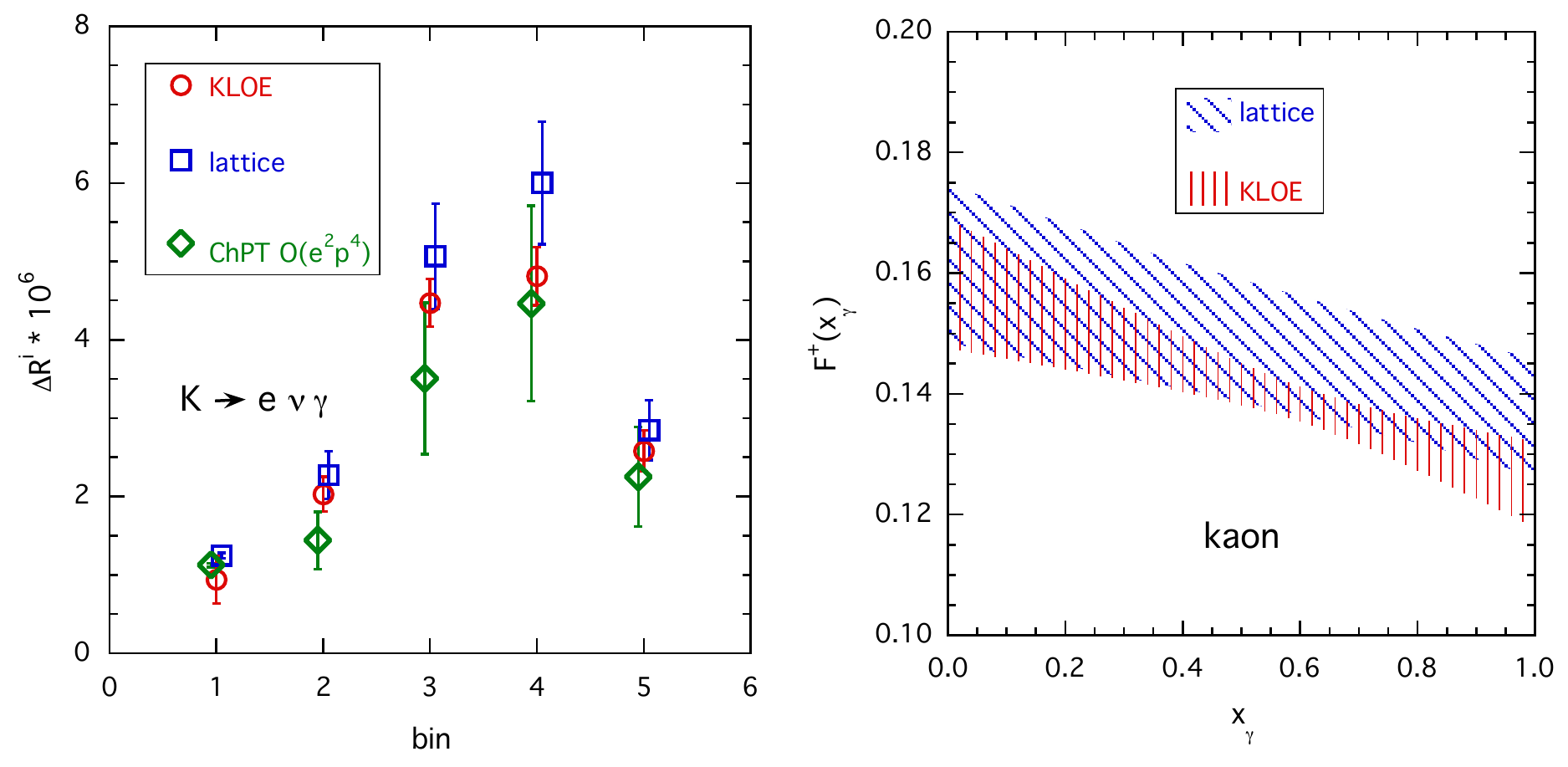}
\end{center}
\vspace{-0.85cm}
\caption{\it \small Left panel: comparison of the KLOE experimental data $\Delta R^{\mathrm{exp}, i}$\,\cite{Ambrosino:2009aa} (red circles) with the theoretical predictions $\Delta R^{\mathrm{th}, i}$, (blue squares) evaluated with the vector and axial form factors of Ref.\,\cite{Desiderio:2020oej} given in Eqs.\,(\ref{eq:FVA_linear})\,-\,(\ref{eq:FA_kaon}), for the 5 bins (see Table\,\ref{tab:KLOE}). The green diamonds correspond to the prediction of ChPT at order ${\cal{O}}(e^2p^4)$, based on the vector and axial form factors given in Eq.\,(\ref{eq:ChPT}).
Right panel: Comparison of the form-factor $F^+(x_\gamma)$ extracted by the KLOE collaboration in Ref.\,\cite{Ambrosino:2009aa} and the theoretical prediction from Eqs.\,(\ref{eq:FVA_linear})\,-\,(\ref{eq:FA_kaon}). The shaded areas represent uncertainties at the level of 1 standard deviation.}
\label{fig:KLOE}
\end{figure}

Using our form factors (\ref{eq:FVA_linear}) with the parameters given in Eqs.\,(\ref{eq:FV_kaon}) and (\ref{eq:FA_kaon}), the INT contributions $\Delta R^{\mathrm{INT}, i}$ turn out to be totally negligible ($\lesssim 10^{-10}$), while the pt term $\Delta R^{\mathrm{pt}, i}$ only contributes significantly in the first bin ($10\,\mathrm{MeV}<E_\gamma<50$\,MeV) where however, it is the dominant contribution leading therefore to a precise prediction for this bin.
For the remaining 4 bins, i.e.~for $i>1$, our theoretical predictions $\Delta R^{\mathrm{th}, i}$ are largely dominated by the SD term, $\Delta R^{\mathrm{SD}, i}$, more precisely by the SD$^+$ contribution related to the square of the form factor $F^+(x_\gamma)$.
Our results are collected in Table\,\ref{tab:KLOE} and shown in the left-hand plot in Fig.\,\ref{fig:KLOE} together with the experimental data $\Delta R^{\mathrm{exp}, i}$ from KLOE.
For all bins a consistency between theory and experiment is observed within about $1$ standard deviation. 
This consistency is underlined in the right-hand plot of Fig.\,\ref{fig:KLOE}, where we compare the form-factor $F^+(x_\gamma)$ extracted by the KLOE collaboration in Ref.\,\cite{Ambrosino:2009aa} with our theoretical prediction. 

\begin{table}[!hbt]
\renewcommand{\arraystretch}{1.2}	 
\begin{center}	
{\footnotesize
\begin{tabular}{||c|c|c||c||c|c|c||c||}
\hline
bin & $E_\gamma$ (MeV) & $p_e$ (MeV) & $\Delta R^{\mathrm{exp}, i} \times 10^6$ & $\Delta R^{\mathrm{SD}, i} \times 10^6$ & $\Delta R^{\mathrm{th}, i} \times 10^6$ & exp / th & ChPT\\
\hline
1 & 10 - 50      & $> 200$ & $~0.94 \pm 0.30 \pm 0.03$ & $~0.26 \pm 0.04$ & $~1.25 \pm 0.04$ & $0.75 \pm 0.24$ & $1.13 \pm 0.03$\\
2 & 50 - 100    & $> 200$ & $~2.03 \pm 0.22 \pm 0.02$ & $~2.26 \pm 0.30$ & $~2.28 \pm 0.30$ & $0.89 \pm 0.15$ & $1.44 \pm 0.36$\\
3 & 100 - 150  & $> 200$ & $~4.47 \pm 0.30 \pm 0.03$ & $~5.06 \pm 0.67$ & $~5.07 \pm 0.67$ & $0.88 \pm 0.13$ & $3.50 \pm 0.96$\\
4 & 150 - 200  & $> 200$ & $~4.81 \pm 0.37 \pm 0.04$ & $~6.00 \pm 0.78$ & $~6.00 \pm 0.78$ & $0.80 \pm 0.12$ & $4.46 \pm 1.25$\\
5 & 200 - 250  & $> 200$ & $~2.58 \pm 0.26 \pm 0.03$ & $~2.85 \pm 0.38$ & $~2.85 \pm 0.38$ & $0.91 \pm 0.15$ & $2.25 \pm 0.63$\\
\hline
1-5 & 10 - 250 & $> 200$ & $14.83 \pm 0.66 \pm 0.13$ & $16.43 \pm 2.12$ & $17.43 \pm 2.12$ & $0.85 \pm 0.11$ & $12.79 \pm 3.24$\\
\hline
\end{tabular}
}
\end{center}
\renewcommand{\arraystretch}{1.0}
\caption{\it \small Values of the KLOE experimental data $\Delta R^{\mathrm{exp}, i}$\,\cite{Ambrosino:2009aa} and of the theoretical predictions $\Delta R^{\mathrm{SD}, i}$ and $\Delta R^{\mathrm{th}, i}$, evaluated with the vector and axial form factors of Ref.\,\cite{Desiderio:2020oej} given in Eqs.\,(\ref{eq:FVA_linear})-(\ref{eq:FA_kaon}), tabulated in the 5 bins of the photon's energy adopted by the KLOE experiment on $K \to e \nu \gamma$ decays. The seventh column is the ratio between the experimental data and our theoretical  predictions. In the fourth column the first error is statistical and the second one is systematic. The last column shows the prediction of ChPT at order ${\cal{O}}(e^2p^4)$, based on the vector and axial form factors given in Eq.\,(\ref{eq:ChPT}).}
\label{tab:KLOE}
\end{table} 

In order to provide a more quantitative measure of the overall level of agreement between theory and experiment we evaluate the reduced $\chi^2$-variable, defined as
\be
    \label{eq:chi2_reduced}
    \chi_{\mathrm{red}}^2 \equiv \frac{1}{N_{\mathrm{bins}}} \sum_{i, j = 1}^{N_{\mathrm{bins}}} \left( \Delta R^{\mathrm{exp},i} - \Delta R^{\mathrm{th},i} \right) C_{ij}^{-1} \left( \Delta R^{\mathrm{exp},j} - \Delta R^{\mathrm{th},j} \right) ~ , ~
\ee
where $C^{-1}$ is the inverse of the global (experiment+theory) covariance matrix and $N_{bins}$ is the number of data points.
Since for all the experiments the covariance matrix is unavailable, we include only the correlations of the theoretical predictions.
For the comparison of the theoretical predictions with the results from the KLOE experiment we find $\chi_{\mathrm{red}}^2 \simeq 0.7$.

In Table\,\ref{tab:KLOE} the last column contains the predictions of ChPT at order ${\cal{O}}(e^2p^4)$, i.e.\,based on the following vector and axial form factors
\bea
    \label{eq:ChPT}
    && C_V^{\mathrm{ChPT}} = \frac{m_P}{4 \pi^2 f_P} ~, \qquad \qquad \qquad D_V^{\mathrm{ChPT}} = 0 ~ , ~ \nonumber \\[2mm]
    && C_A^{\mathrm{ChPT}} = \frac{8 m_P}{f_P} (L_9^r + L_{10}^r)~, \qquad D_A^{\mathrm{ChPT}} = 0 ~ 
\eea
with $L_9^r + L_{10}^r = 0.0017\,(7)$\,\cite{Bijnens:2014lea} and taking $m_K / f_K = 493.7\,{\rm MeV} / 156.1\,{\rm MeV}$.
These predictions are in good agreement with the experimental points to within about $1$ standard deviation ($\chi_{\mathrm{red}}^2 \simeq 1.3$).

\section{Comparison with the E787, ISTRA+ and OKA experiments}
\label{sec:kaon}

In this Section we compare our lattice predictions with the experimental data on the leptonic radiative decays of kaons into muons, $K_{\mu2\gamma}$, obtained by the E787\,\cite{Adler:2000vk}, ISTRA+\,\cite{Duk:2010bs} and OKA\,\cite{Kravtsov:2019amb} collaborations.
The kinematical regions in terms of photon and lepton energies were suitably chosen in order to enhance the contributions of the SD$^+$ term in the case of the E787 experiment and of the INT$^-$ term in the case of the ISTRA+ and OKA experiments.
We remind the reader that the SD$^+$ and INT$^-$ terms are related to the square of the form factor $F^+$ and to the form factor $F^-$, respectively.

\subsection{The E787 experiment}
\label{sec:E787}

In Ref.\,\cite{Adler:2000vk} the E787 Collaboration has investigated the $K_{\mu2\gamma}$ decay for photon energies in the range $90\,{\rm MeV} < E_\gamma < E_\gamma^{max} \simeq 235\,{\rm MeV}$ with the constraint that the muon kinetic energy is larger than 137\,MeV (i.e.~$E_\mu > m_\mu + 137\,\mbox{MeV} \simeq 243$\,MeV).
In such kinematical regions the radiated photons come mainly from the pt contribution and the SD$^+$ terms\,\cite{Adler:2000vk}.
In order to compare their results with those from other experiments, the E787 data are integrated over the small allowed range of muon energies $243$\,MeV $< E_\mu \leq E_\mu^{max} \simeq 258$\,MeV, assuming a constant acceptance, to obtain
the differential branching ratio
\be
    \label{eq:BR_theta}
    \frac{dR^{\mathrm{exp}}}{d \cos(\theta_{\mu \gamma})} = \frac{1}{\Gamma(K _{\mu2[\gamma]})} \left[ \frac{d\Gamma(K _{\mu2\gamma})}
        {d \cos(\theta_{\mu \gamma})} \right]_{E_\gamma > 90\, {\rm MeV}, E_\mu > 243\, {\rm MeV}}
\ee
as a function of the emission angle $\theta_{\mu \gamma}$ between the muon and the photon in the kaon rest-frame.

At leading order, ${\cal{O}}(\alpha_{\mathrm{em}})$, the theoretical prediction for $dR^{\mathrm{th}} / d \cos(\theta_{\mu \gamma})$ can be written as the sum of the following five terms
\be
    \label{eq:BR_dec}
    \frac{dR^{\mathrm{th}}}{d \cos(\theta_{\mu \gamma})} = \frac{dR^{\mathrm{pt}}}{d \cos(\theta_{\mu \gamma})} + \frac{dR^{\mathrm{SD}^+}}{d \cos(\theta_{\mu \gamma})} + 
        \frac{dR^{\mathrm{INT}^+}}{d \cos(\theta_{\mu \gamma})}+
        \frac{dR^{\mathrm{SD}^-}}{d \cos(\theta_{\mu \gamma})}+
\frac{dR^{\mathrm{INT}^-}}{d \cos(\theta_{\mu \gamma})}\,,
\ee
where
\bea
   \label{eq:BRs}
    \frac{dR^{\mathrm{pt\,(SD^\pm,INT^\pm)}}}{d \cos(\theta_{\mu \gamma})} & = & \int_{x_\gamma^{min}}^{1 - r_\mu^2} d x_\gamma ~ 
        \int_{x_\mu^{min}}^1 d x_\mu \left[ \frac{d^2 R_1^{\mathrm{pt\,(SD^\pm,INT^\pm)}}}{d x_\gamma dx_\mu} \right]  \nonumber \\[2mm]
        & \times & \delta\left[ \cos(\theta_{\mu \gamma}) - \frac{x_\mu + r_\mu^2 - 2(x_\mu + x_\gamma - 1) / x_\gamma}
                        {\sqrt{(x_\mu + r_\mu^2)^2 - 4 r_\mu^2}}\right]\,,
\eea
with $x_\gamma^{min} = [2 (90\,{\rm MeV}) / m_K] \simeq 0.36$, $x_\mu^{min} = [2 (243\,{\rm MeV}) / m_K - r_\mu^2] \simeq 0.94$ and $r_\mu = m_\mu^2 / m_K^2 \simeq 0.046$, while the double differential branching ratios $d^2 R_1^{\mathrm{pt}(SD^\pm,INT^\pm)} / d x_\gamma dx_\mu$ are given by Eqs.\,(\ref{eq:d2Gamma_pt})\,-\,(\ref{eq:d2Gamma_INT}).
On the right-hand side of Eq.\,(\ref{eq:BR_dec}) the first term is the pt contribution, the second and third terms depend on the form factor $F^+(x_\gamma)$, while the fourth and fifth terms depend on $F^-(x_\gamma)$.

\begin{figure}[htb!]
\begin{center}
\includegraphics[scale=0.80]{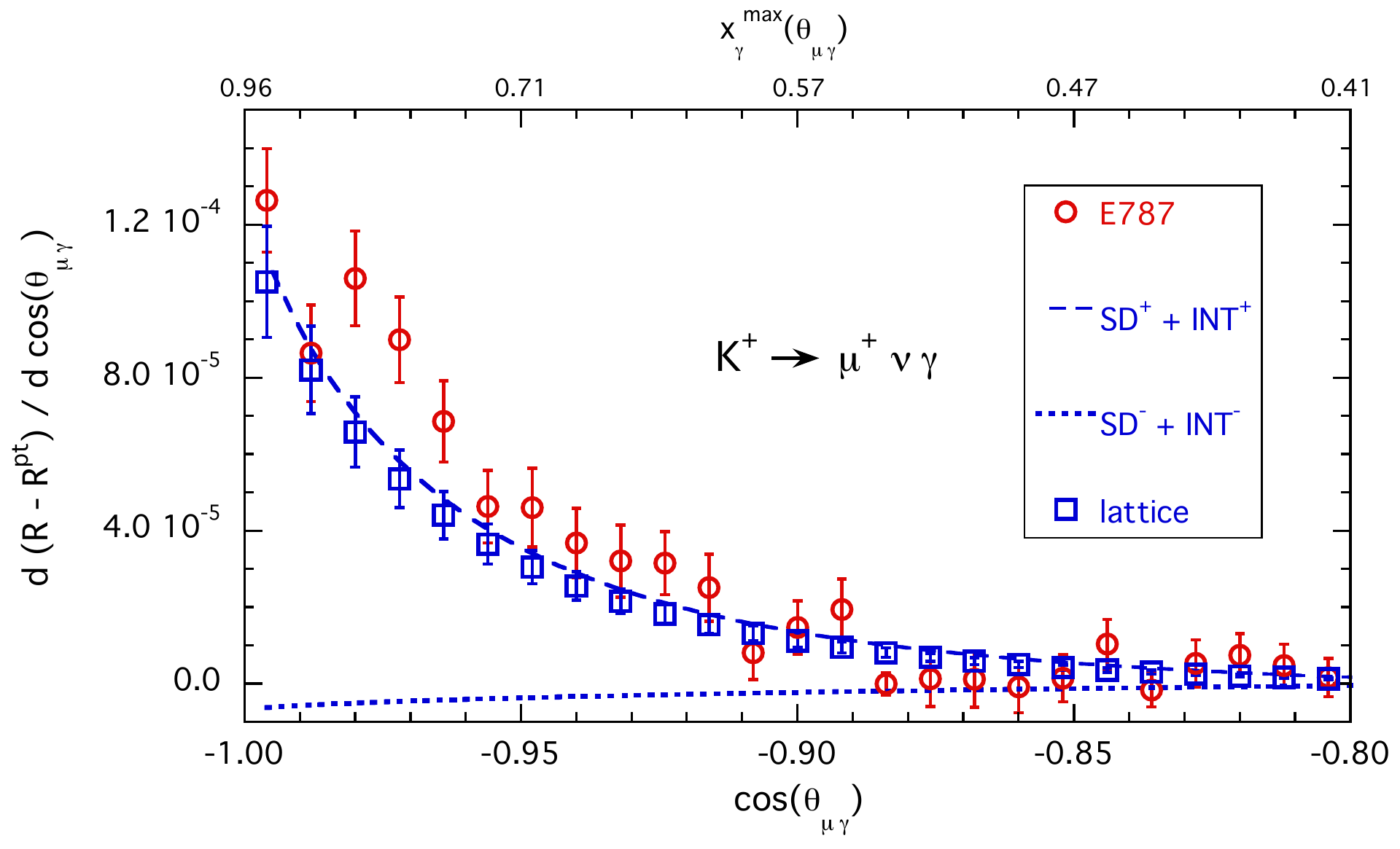}
\end{center}
\vspace{-0.80cm}
\caption{\it \small Comparison of the E787 experimental data after the pt contribution has been subtracted, $d (R^{\mathrm{exp}}-R^{\mathrm{pt}})/ d \cos(\theta_{\mu \gamma})$ (red circles)\,\cite{Adler:2000vk}, with the theoretical predictions $d (R^{\mathrm{th}}-R^{\mathrm{pt}})/ d \cos(\theta_{\mu \gamma})$ (blue squares), evaluated using the lattice form factors of Ref.\,\cite{Desiderio:2020oej} given in Eqs.\,(\ref{eq:FVA_linear})-(\ref{eq:FA_kaon}). The dashed and dotted lines correspond to the contributions $d (R^{\mathrm{SD}^+}+R^{\mathrm{INT^+}}) / d \cos(\theta_{\mu \gamma})$ and 
$d (R^{\mathrm{SD}^-}+R^{\mathrm{INT^-}}) / d \cos(\theta_{\mu \gamma})$
respectively. The upper horizontal axis shows the maximum value of $x_\gamma$, $x_\gamma^{max}(\theta_{\mu \gamma})$, allowed by the value of the angle $\theta_{\mu \gamma}$ taking into account the kinematical cuts of the E787 experiment (see Eq.\,(\ref{eq:BRs})).}
\label{fig:E787}
\end{figure}

Since the pt contribution is a purely kinematical factor, it can be subtracted from the experimental data without introducing any uncertainty. 
The corresponding {\it subtracted} data are compared with our theoretical predictions in Table\,\ref{tab:E787} and in Fig.\,\ref{fig:E787}.
A reasonable agreement is found except for some points at large backward angles, i.e.~at large photon energies, where the tension reaches about 2\,-\,3 standard deviations.
There the data are dominated by the contributions coming from the form factor $F^+(x_\gamma)$.
For the global reduced $\chi^2$-variable (see Eq.\,(\ref{eq:chi2_reduced})) we get $\chi_{\mathrm{red}}^2 \simeq 1.6$.

\begin{table}[!hbt]	 
\begin{minipage}[t]{0.475\linewidth}
{\small
\begin{tabular}{||c||c||c||}
\hline
\rule[-.25cm]{0pt}{20pt}$\mbox{cos}(\theta_{\mu \gamma})$ & ~$\frac{d (R^{\mathrm{exp}}-R^{\mathrm{pt}})}{d \cos(\theta_{\mu \gamma})} \cdot 10^4$~ & ~$\frac{d (R^{\mathrm{th}}-R^{\mathrm{pt}})}{d \cos(\theta_{\mu \gamma})} \cdot 10^4$~\\
\hline
$-0.996$ & $1.264~(135)$ & $1.051~(146)$ \\
$-0.988$ & $0.865~(127)$ & $0.820~(114)$ \\
$-0.980$ & $1.059~(124)$ & $0.658~~(92)$ \\
$-0.972$ & $0.900~(112)$ & $0.536~~(75)$ \\
$-0.964$ & $0.685~(106)$ & $0.440~~(62)$ \\
$-0.956$ & $0.463~~(94)$ & $0.365~~(52)$ \\
$-0.948$ & $0.460~(103)$ & $0.304~~(44)$ \\
$-0.940$ & $0.368~~(91)$ & $0.255~~(37)$ \\
$-0.932$ & $0.320~~(94)$ & $0.215~~(31)$ \\
$-0.924$ & $0.315~~(82)$ & $0.182~~(27)$ \\
$-0.916$ & $0.251~~(88)$ & $0.154~~(23)$ \\
$-0.908$ & $0.081~~(71)$ & $0.131~~(20)$ \\
$-0.900$ & $0.146~~(71)$ & $0.112~~(17)$ \\
\hline
\end{tabular}
}
\end{minipage}
\hfill
\begin{minipage}[t]{0.475\linewidth}
{\small
\begin{tabular}{||c||c||c||}
\hline
\rule[-.25cm]{0pt}{20pt}$\mbox{cos}(\theta_{\mu \gamma})$ & ~$\frac{d (R^{\mathrm{exp}}-R^{\mathrm{pt}})}{d \cos(\theta_{\mu \gamma})} \cdot 10^4$~ & ~$\frac{d (R^{\mathrm{th}}-R^{\mathrm{pt}})}{d \cos(\theta_{\mu \gamma})} \cdot 10^4$~\\
\hline
$-0.892$ & $~~0.194~(79)$ & $0.095~(15)$ \\
$-0.884$ & $-0.001~(28)$ & $0.081~(13)$ \\
$-0.876$ & $~~0.013~(74)$ & $0.069~(11)$ \\
$-0.868$ & $~~0.011~(74)$ & $0.059~~(9)$ \\
$-0.860$ & $-0.009~(68)$ & $0.050~~(8)$ \\
$-0.852$ & $~~0.014~(62)$ & $0.042~~(7)$ \\
$-0.844$ & $~~0.104~(65)$ & $0.036~~(6)$ \\
$-0.836$ & $-0.017~(44)$ & $0.030~~(5)$ \\
$-0.828$ & $~~0.053~(62)$ & $0.025~~(4)$ \\
$-0.820$ & $~~0.074~(56)$ & $0.020~~(3)$ \\
$-0.812$ & $~~0.047~(56)$ & $0.016~~(3)$ \\
$-0.804$ & $~~0.016~(50)$ & $0.013~~(2)$ \\
&& \\
\hline
\end{tabular}
}
\end{minipage}
\vspace{0.25cm}
\caption{\it \small Results from the E787 experiment\,\cite{Adler:2000vk} (see text) after subtraction of the pt contribution, $d (R^{\mathrm{exp}}-R^{\mathrm{pt}})/ d \cos(\theta_{\mu \gamma})$ for selected values of $\cos\theta_{\mu \gamma}$, together with our theoretical predictions $d (R^{\mathrm{th}}-R^{\mathrm{pt}})/ d \cos(\theta_{\mu \gamma})$
evaluated using the vector and axial form factors of Ref.\,\cite{Desiderio:2020oej} given in Eqs.\,(\ref{eq:FVA_linear})\,-\,(\ref{eq:FA_kaon}).}
\label{tab:E787}
\end{table} 

Note that, though generally small, the relative contribution of SD$^- + $INT$^-$, which depends on the form factor $F^-(x_\gamma)$, becomes more important as $\cos(\theta_{\mu \gamma})$ increases (i.e.~as $x_\gamma$ decreases), reaching about 20\,-\,30\% of the term SD$^+ + $INT$^+$ at the lowest available values of $x_\gamma$. 

We remind the reader that, as shown in Sec.\,\ref{sec:KLOE}, our lattice form factor $F^+(x_\gamma)$ leads to a good description of the KLOE data\,\cite{Ambrosino:2009aa}. A consequence of this is that the tension between our theoretical predictions and the E787 data which is visible at large $x_\gamma$ in Fig.\,\ref{fig:E787} is not unexpected because of a tension between the two experiments.
The KLOE collaboration has estimated $F^+(x_\gamma =1)$ to be equal to $0.125 \pm 0.007_{\mathrm{\scriptsize stat}} \pm 0.001_{\mathrm{\scriptsize syst}}$\,\cite{Ambrosino:2009aa}, while the estimate of E787, assuming a constant form factor, is $0.165 \pm 0.007_{\mathrm{\scriptsize stat}} \pm 0.011_{\mathrm{\scriptsize syst}}$\,\cite{Adler:2000vk}.
The difference is at the level of about $3$ standard deviations (see also the discussion in Sec.\,\ref{sec:SM_fit} below).
Our theoretical prediction for this quantity is $F^+(x_\gamma =1)=0.1362 \pm 0.0096$.

\begin{figure}[htb!]
\begin{center}
\includegraphics[scale=0.80]{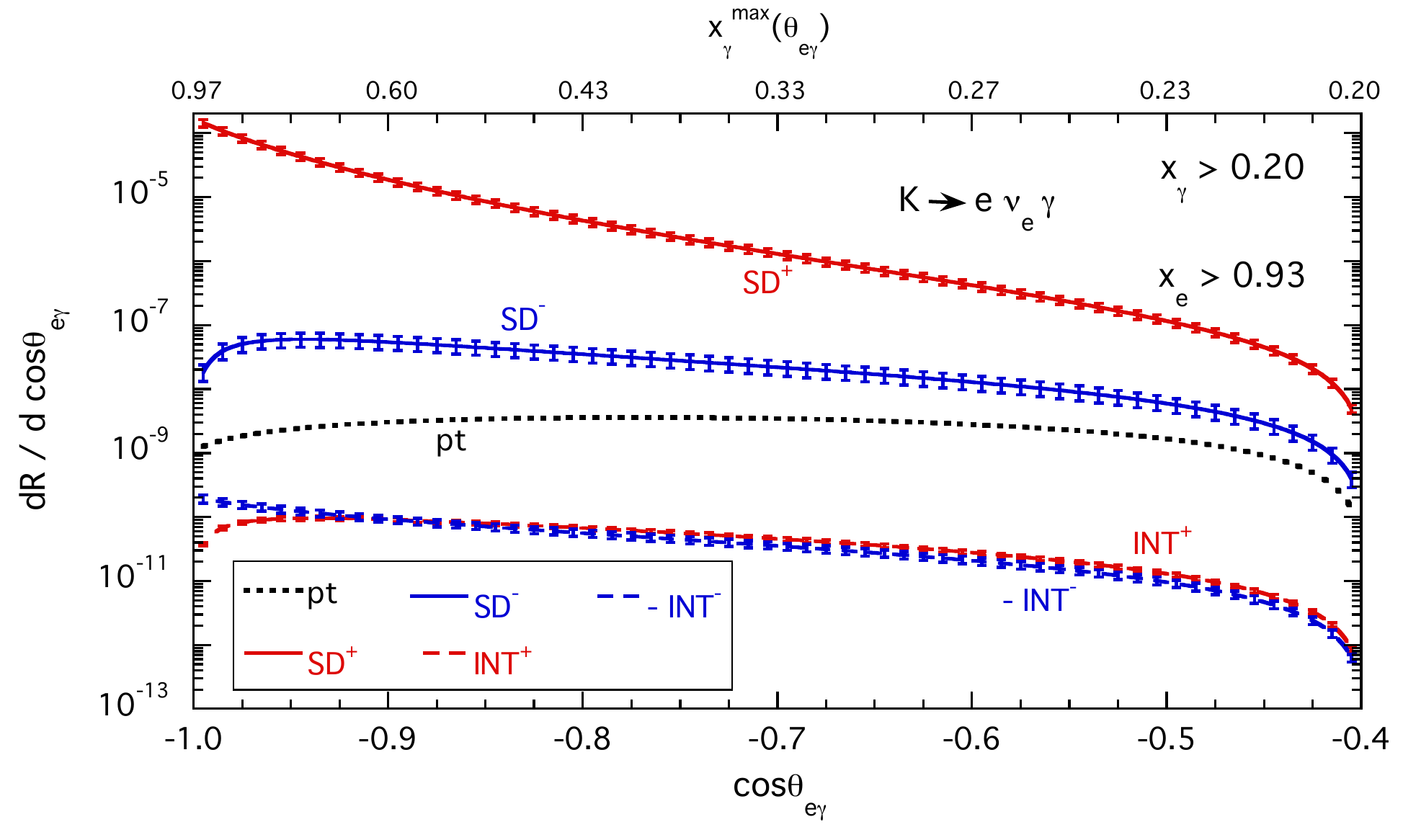}
\end{center}
\vspace{-0.80cm}
\caption{\it \small Results for the pt, SD$^+$, SD$^-$, INT$^+$ and INT$^-$ contributions to the differential branching ratio (\ref{eq:NA62}) as a function of the emission angle $\theta_{e \gamma}$ for the decay process $K_{e2\gamma}$, calculated using the lattice form factors of Ref.\,\cite{Desiderio:2020oej}, given in Eqs.\,(\ref{eq:FVA_linear})-(\ref{eq:FA_kaon}), with the kinematical cuts $x_\gamma > 0.2$ ($E_\gamma > 49$\,{\rm MeV}) and $x_e > 0.93$ ($E_e > 230$\,{\rm MeV}).}
\label{fig:NA62}
\end{figure}

Thus, further experimental investigations of the form factor $F^+(x_\gamma)$ in radiative kaon decays into electrons and muons are required. 
In particular, an investigation of the decay $K_{e2\gamma}$ at large electron energies will provide the opportunity for an accurate determination of $|F^+(x_\gamma)|$ for a wide range of values of $x_\gamma$.
This is illustrated in Fig.\,\ref{fig:NA62}, where the pt, SD$^+$, SD$^-$, INT$^+$ and INT$^-$ contributions to the differential branching ratio
\be
    \label{eq:NA62}
    \frac{dR}{d \cos(\theta_{e \gamma})} = \frac{1}{\Gamma(K _{e2[\gamma]})} \left[ \frac{d\Gamma(K _{e2\gamma})}
        {d\cos(\theta_{e \gamma})} \right]_{x_\gamma > 0.2,\, x_e > 0.93}
\ee
are shown as a function of the emission angle $\theta_{e \gamma}$ between the electron and the photon (in the kaon rest-frame) after considering the kinematical cuts $x_\gamma > 0.2$ ($E_\gamma > 49$ MeV) and $x_e > 0.93$ ($E_e > 230$ MeV).
These kinematical cuts are indicative of a possible definition of a signal region with minimal background contamination both from the pt contribution to $K_{e2\gamma}$ and from the semileptonic $K_{e3}$ process in a fixed-target forward detector such as that in the NA62 experiment\,\cite{NA62}\,\footnote{We thank members of the NA62 experiment for discussions on this point.}.

\subsection{The ISTRA+ and OKA experiments}
\label{sec:OKA}

In Refs.\,\cite{Duk:2010bs} and\,\cite{Kravtsov:2019amb} the ISTRA+ and OKA collaborations have selected appropriate kinematical regions (strips) in order to determine the contribution of the interference term INT$^-$.
For each strip, specific bins are selected in the photon and muon variables $x_\gamma$ and $y_\mu \equiv 2 E_\mu / m_K = x_\mu + r_\mu^2$, where $E_\mu$ is the muon energy in the kaon rest frame. 
A further constraint $\cos(\theta_{\mu \gamma}) > \cos(\theta_{\mathrm{cut}})$ is imposed on the emission angle $\theta_{\mu \gamma}$ between the muon and the photon. 
The kinematical cuts are collected in Tables\,\ref{tab:ISTRA} and~\ref{tab:OKA} and can be taken into account by using the kinematical functions $\widetilde{f}_{\mathrm{pt,SD,INT}}(x_\gamma; x_0, x_1)$, given in Eqs.\,(\ref{eq:ftildePT})-(\ref{eq:ftildeINT-}), with
\bea
     \label{eq:x0_mu}
     x_0 & = & {\rm max}\left[ y_\mu^{min}(i) - r_\mu^2, 1 - x_\gamma + r_\mu^2 \frac{x_\gamma}{1 - x_\gamma} \right] ~ , ~ \\[2mm]
     \label{eq:x1_mu}
     x_1   & = & {\rm min}\left[ x_+^i, y_\mu^{max}(i) - r_\mu^2, 1 \right] ~ , ~
\eea
where the index $i$ labels the strip and $x_+$ is equal to
\be
     \label{eq:xcutp}
    x_+ = \frac{2}{a} \left[ b + \sqrt{b^2 - a\,c}\, \right] -r_\mu^2
\ee
with
\bea
      \label{eq:a}
      a & = & \left( \frac{2}{x_\gamma} - 1 \right)^2 - \cos^2(\theta_{\mathrm{cut}})\,, \\[2mm]
      \label{eq:b}
      b & = & \left( \frac{2}{x_\gamma} - 1 \right) \frac{1 - x_\gamma + r_\mu^2}{x_\gamma}\,,\\[2mm]
      \label{eq:c}
      c & = & r_\mu^2\cos^2(\theta_{\mathrm{cut}}) + \left( \frac{1 - x_\gamma + r_\mu^2}{x_\gamma} \right)^2 
\eea
and $\cos(\theta_{\mathrm{cut}})$ given in Tables\,\ref{tab:ISTRA} and~\ref{tab:OKA} for each strip.

\begin{table}[!hbt]	 
\begin{minipage}[t]{0.475\linewidth}
{\footnotesize
\begin{tabular}{||c||c|c|c||}
\hline
strip & $x_\gamma$ & $y_\mu$ & $ \cos(\theta_{\mathrm{cut}})$\\
\hline
01 & $0.05 < x_\gamma < 0.10$ & $0.90 < y_\mu < 1.10$ & $- 0.8$ \\
02 & $0.10 < x_\gamma < 0.15$ & $0.90 < y_\mu < 1.10$ & $- 0.8$ \\
03 & $0.15 < x_\gamma < 0.20$ & $0.85 < y_\mu < 1.00$ & $- 0.8$ \\
04 & $0.20 < x_\gamma < 0.25$ & $0.80 < y_\mu < 0.95$ & $- 0.2$ \\
05 & $0.25 < x_\gamma < 0.30$ & $0.75 < y_\mu < 0.90$ & $- 0.3$ \\
06 & $0.30 < x_\gamma < 0.35$ & $0.72 < y_\mu < 0.87$ & $- 0.4$ \\
07 & $0.35 < x_\gamma < 0.40$ & $0.65 < y_\mu < 0.85$ & $- 0.3$ \\
08 & $0.40 < x_\gamma < 0.45$ & $0.62 < y_\mu < 0.85$ & $- 0.5$ \\
09 & $0.45 < x_\gamma < 0.50$ & $0.57 < y_\mu < 0.80$ & $- 0.7$ \\
10 & $0.50 < x_\gamma < 0.55$ & $0.52 < y_\mu < 0.75$ & $- 1.0$ \\
11 & $0.55 < x_\gamma < 0.60$ & $0.48 < y_\mu < 0.70$ & $- 1.0$ \\
\hline
\end{tabular}
}
\vspace{0.25cm}
\caption{\it \small Kinematical cuts adopted in the ISTRA+ experiment of Ref.\,\cite{Duk:2010bs} (see text). }
\label{tab:ISTRA}
\end{minipage}
\hfill
\begin{minipage}[t]{0.475\linewidth}
{\footnotesize
\begin{tabular}{||c||c|c|c||}
\hline
strip & $x_\gamma$ & $y_\mu$ & $ \cos(\theta_{\mathrm{cut}})$\\
\hline
01 & $0.10 < x_\gamma < 0.15$ & $0.89 < y_\mu < 1.01$ & $- 0.8$ \\
02 & $0.15 < x_\gamma < 0.20$ & $0.85 < y_\mu < 1.01$ & $- 0.2$ \\
03 & $0.20 < x_\gamma < 0.25$ & $0.80 < y_\mu < 1.00$ & $- 0.2$ \\
04 & $0.25 < x_\gamma < 0.30$ & $0.75 < y_\mu < 0.97$ & $- 0.4$ \\
05 & $0.30 < x_\gamma < 0.35$ & $0.70 < y_\mu < 0.93$ & $- 0.4$ \\
06 & $0.35 < x_\gamma < 0.40$ & $0.66 < y_\mu < 0.90$ & $- 0.5$ \\
07 & $0.40 < x_\gamma < 0.45$ & $0.62 < y_\mu < 0.88$ & $- 0.5$ \\
08 & $0.45 < x_\gamma < 0.50$ & $0.58 < y_\mu < 0.86$ & $- 0.6$ \\
09 & $0.50 < x_\gamma < 0.55$ & $0.54 < y_\mu < 0.83$ & $- 0.6$ \\
10 & $0.55 < x_\gamma < 0.60$ & $0.50 < y_\mu < 0.80$ & $- 0.6$ \\
&&&\\
\hline
\end{tabular}
}
\vspace{0.25cm}
\caption{\it \small The same as in Table~\ref{tab:ISTRA}, but in the case of the OKA experiment of Ref.\,\cite{Kravtsov:2019amb}.}
\label{tab:OKA}
\end{minipage}
\end{table} 

In both experiments the measured observable is the ratio $N_{\mathrm{exp}} / N_{\mathrm{pt}}$ of the number of observed photons in each strip to the number of pt (or inner-bremsstrahlung) events. $N_{\mathrm{pt}}$ is estimated using the Geant3 package\,\cite{Geant3}.

\begin{figure}[htb!]
\begin{center}
\includegraphics[scale=0.85]{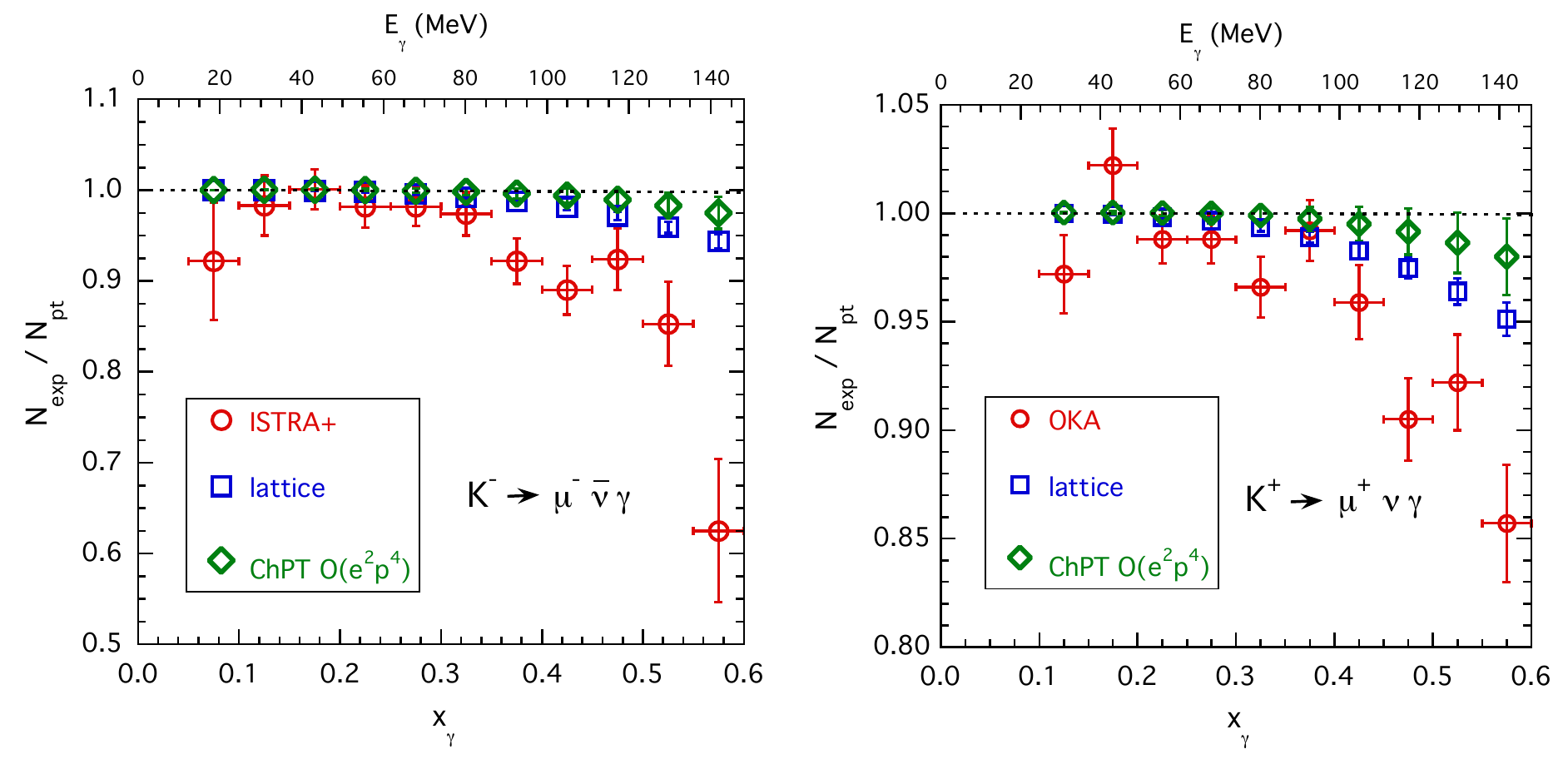}
\end{center}
\vspace{-0.85cm}
\caption{\it \small Comparison of the experimental results from the ISTRA+\,\cite{Duk:2010bs} (left panel) and OKA\,\cite{Kravtsov:2019amb} (right panel) collaborations with our theoretical predictions. The predictions were evaluated using the vector and axial form factors of Ref.\,\cite{Desiderio:2020oej}, given in Eqs.\,(\ref{eq:FVA_linear})\,-\,(\ref{eq:FA_kaon}), for the kinematical strips selected by the two experiments on $K_{\mu2\gamma}$ decays. The green diamonds correspond to the prediction of ChPT at order ${\cal{O}}(e^2p^4)$, based on the vector and axial form factors given in Eq.\,(\ref{eq:ChPT}). Note the different scales of the vertical axes in the two panels.}
\label{fig:ISTRA+OKA}
\end{figure}

The comparison of the experimental results with our predictions, and also with those obtained using ChPT at order ${\cal{O}}(e^2p^4)$ based on the vector and axial-vector form factors given in Eq.\,(\ref{eq:ChPT}) with $m_K / f_K = 493.7~{\rm MeV} / 156.1~{\rm MeV}$, is presented in Table\,\ref{tab:ISTRA+OKA} and in Fig.\,\ref{fig:ISTRA+OKA}.
It can clearly be seen that at large photon energies there is a significant tension between the experimental data and our non-perturbative results (and also those obtained using ChPT).
For the global reduced $\chi^2$-variable (see Eq.\,\ref{eq:chi2_reduced})) we find $\chi_{\mathrm{red}}^2 \simeq 3.9$ and $\simeq 3.4$ for the ISTRA+ and OKA experiments respectively.
Thus, improved determinations of the form factor $F^-(x_\gamma)$ are required from both experiment and theory in order to consolidate or eliminate the discrepancies. 

\begin{table}[!hbt]	 
\begin{minipage}[t]{0.475\linewidth}
{\footnotesize
\begin{tabular}{||c||c||c|c||}
\hline
strip & ~~$N_{\mathrm{exp}} / N_{\mathrm{pt}}$~~& ~~$N_{\mathrm{th}} / N_{\mathrm{pt}}$~~ & ~~~~ChPT~~~~\\
\hline
01 & $0.922~(65)$ & $1.0001~~(1)$ & $1.0002~~~(1)$ \\
02 & $0.983~(33)$ & $1.0001~~(2)$ & $1.0004~~~(4)$ \\
03 & $1.001~(22)$ & $0.9996~~(4)$ & $1.0005~~~(8)$ \\
04 & $0.982~(23)$ & $0.9983~~(7)$ & $1.0002~~(14)$ \\
05 & $0.982~(21)$ & $0.9956~(11)$ & $0.9994~~(23)$ \\
06 & $0.974~(24)$ & $0.9922~(17)$ & $0.9981~~(36)$ \\
07 & $0.922~(25)$ & $0.9873~(25)$ & $0.9963~~(54)$ \\
08 & $0.890~(27)$ & $0.9816~(35)$ & $0.9942~~(77)$ \\
09 & $0.924~(34)$ & $0.9718~(47)$ & $0.9895~(104)$ \\
10 & $0.853~(46)$ & $0.9591~(62)$ & $0.9830~(137)$ \\
11 & $0.625~(79)$ & $0.9436~(81)$ & $0.9747~(176)$ \\
\hline
\end{tabular}
}
\end{minipage}
\hfill
\begin{minipage}[t]{0.475\linewidth}
{\footnotesize
\begin{tabular}{||c||c||c|c||}
\hline
strip & ~~$N_{\mathrm{exp}} / N_{\mathrm{pt}}$~~& ~~$N_{\mathrm{th}} / N_{\mathrm{pt}}$~~ & ~~~~ChPT~~~~\\
\hline
01 & $0.972~(18)$ & $1.0000~~(2)$ & $1.0004~~~(3)$ \\
02 & $1.022~(17)$ & $0.9995~~(3)$ & $1.0004~~~(7)$ \\
03 & $0.988~(11)$ & $0.9983~~(7)$ & $1.0002~~(14)$ \\
04 & $0.988~(11)$ & $0.9966~(11)$ & $1.0001~~(24)$ \\
05 & $0.966~(14)$ & $0.9935~(17)$ & $0.9991~~(38)$ \\
06 & $0.992~(14)$ & $0.9889~(25)$ & $0.9975~~(56)$ \\
07 & $0.959~(17)$ & $0.9827~(35)$ & $0.9950~~(79)$ \\
08 & $0.905~(19)$ & $0.9747~(47)$ & $0.9916~(107)$ \\
09 & $0.922~(22)$ & $0.9641~(61)$ & $0.9865~(139)$ \\
10 & $0.857~(27)$ & $0.9512~(78)$ & $0.9800~(177)$ \\
&&&\\
\hline
\end{tabular}
}
\end{minipage}
\vspace{0.25cm}
\caption{\it \small Values of $N_{\mathrm{exp}}/N_{\mathrm{pt}}$ (see text) for the ISTRA+\,\cite{Duk:2010bs} (left panel) and OKA experiments\,\cite{Kravtsov:2019amb} (right panel), compared to our theoretical predictions $N_{th}/N_{pt}$, evaluated using the vector and axial form factors of Ref.\,\cite{Desiderio:2020oej} given in Eqs.\,(\ref{eq:FVA_linear})\,-\,(\ref{eq:FA_kaon}), for the kinematical strips selected by the two experiments (see Tables\,\ref{tab:ISTRA} and\,\ref{tab:OKA}). The fourth columns correspond to the predictions of ChPT at order ${\cal{O}}(e^2p^4)$, based on the vector and axial form factors given in Eq.\,(\ref{eq:ChPT}).}
\label{tab:ISTRA+OKA}
\end{table} 

\section{Comparison with the experimental data from the PIBETA collaboration}
\label{sec:PIBETA}

In Ref.\,\cite{Bychkov:2008ws} the PIBETA Collaboration has investigated the radiative pion decay into electrons $\pi_{e2\gamma}$ and has measured the following branching ratios
\be
     \label{eq:branching}
     \Delta R^{\mathrm{exp}, i} \equiv \frac{1}{\Gamma(\pi \to \mu \nu[\gamma])} \int_{E_\gamma^i}^{E_\gamma^{max}} d E_\gamma ~ \int_{E_e^i}^{E_e^{max}} d E_e ~
                                           \left[ \frac{d^2\Gamma(\pi ^+ \to e^+ \nu \gamma)}{dE_\gamma dE_e}\right]_{\theta_{e \gamma} > 40^\circ} 
\ee
integrated in four different kinematical regions of photon and electron energies with the constraint $\theta_{e \gamma} > 40^\circ$.
The kinematical regions are labelled as $i = A, B, C, O$ and the values of the minimum photon and electron energies are, respectively, $E_\gamma^i = \{ 50, 50, 10, 10\}$\,MeV and $E_e^i = \{ 50, 10, 50, m_e \}$\,MeV.
The maximum photon and electron energies are $E_\gamma^{max} \simeq E_e^{max} \simeq m_\pi  / 2 \simeq 70$\,MeV.
The region $O$ is a combination of the other three regions supplemented with extrapolations based on Monte Carlo simulations\,\cite{Bychkov:2008ws}.

As was the case for $K_{e2\gamma}$ decays in Eq.\,(\ref{eq:DeltaRdec}), at order ${\cal{O}}(\alpha_{\mathrm{em}})$ the theoretical prediction for each bin for $\pi_{e2\gamma}$ decays, $\Delta R^{\mathrm{th}, i}$, can be decomposed into the sum of three terms
\be
    \label{eq:DeltaBdec}
    \Delta R^{\mathrm{th}, i} = \Delta R^{\mathrm{pt}, i} + \Delta R^{\mathrm{SD}, i} + \Delta R^{\mathrm{INT}, i} ~ , ~
\ee
where in this case
\be
   \label{eq:DeltaBi}
    \Delta R^{\mathrm{pt\,(SD,INT), i}} = \frac{\Gamma^{(0)}(\pi \to e \nu)}{\Gamma^{(0)}(\pi \to \mu \nu)} ~ \int_{2E_\gamma^i/m_\pi}^{1 - r_e^2} d x_\gamma ~ 
                                             \left[ \frac{dR_1^{\mathrm{pt\,(SD,INT)}}}{d x_\gamma} \right]_{E_e > E_e^i,\, \theta_{e \gamma} > 40^\circ}
\ee
with 
\be
    \frac{\Gamma^{(0)}(\pi_{e2})}{\Gamma^{(0)}(\pi_{\mu2})} = \frac{m_e^2}{m_\mu^2} \frac{(1 - r_e^2)^2}{(1 - r_\mu^2)^2} 
                                                                                                  \simeq 1.2834 \times 10^{-4} ~ 
\ee
and $r_e = m_e / m_\pi$ and $r_\mu = m_\mu / m_\pi$.

The constraint on the electron energies $E_e > E_e^i$ implies $x_e > x_{min}^i$, where
\be
     \label{eq:xmin_i}
     x_{min}^i \equiv \frac{2 E_e^i}{m_\pi} - r_e^2 ~  , ~
\ee
while, using momentum conservation, the constraint $\theta_{e \gamma} > \theta_{\mathrm{cut}} = 40^\circ$ implies $x_e > x_-$  for $x_\gamma \leq 1 - r_e$, where 
\be
     \label{eq:xcutm}
    x_- = \frac{2}{a} \left[ b - \sqrt{b^2 - a\, c} \,\right] -r_e^2
\ee
and $a$, $b$ and $c$ are given by Eqs.\,(\ref{eq:a})\,-\,(\ref{eq:c}) (replacing $r_\mu$ with $r_e$).
In the region $1 - r_e < x_\gamma \leq 1 - r_e^2$ the constraint $\theta_{e \gamma} > \theta_{\mathrm{cut}} = 40^\circ$ is always satisfied.

The contributions $[ dR_1^{\mathrm{pt,SD,INT}} / d x_\gamma ]_{E_e > E_e^i, \theta_{e \gamma} > 40^\circ}$ are given by Eqs.\,(\ref{eq:deltaR1_pt_x0})\,-\,(\ref{eq:deltaR1_INT_x0}), with $m_K,\,f_K$ now replaced by $m_\pi,\,f_\pi$, and with $x_0$ equal to
\bea
    x_0 & = & {\rm max}\left(x_{min}^i, x_-, 1 - x_\gamma + r_e^2 \frac{x_\gamma}{1 - x_\gamma} \right) \qquad ~ {\rm for} ~ x_\gamma \leq 1 - r_e ~ , ~ 
                     \nonumber  \\
           & = & {\rm max}\left(x_{min}^i, 1 - x_\gamma + r_e^2 \frac{x_\gamma}{1 - x_\gamma} \right) \qquad \qquad {\rm for} ~ x_\gamma > 1 - r_e ~ . ~
    \label{eq:x0_th}
\eea

\begin{figure}[htb!]
\begin{center}
\includegraphics[scale=0.75]{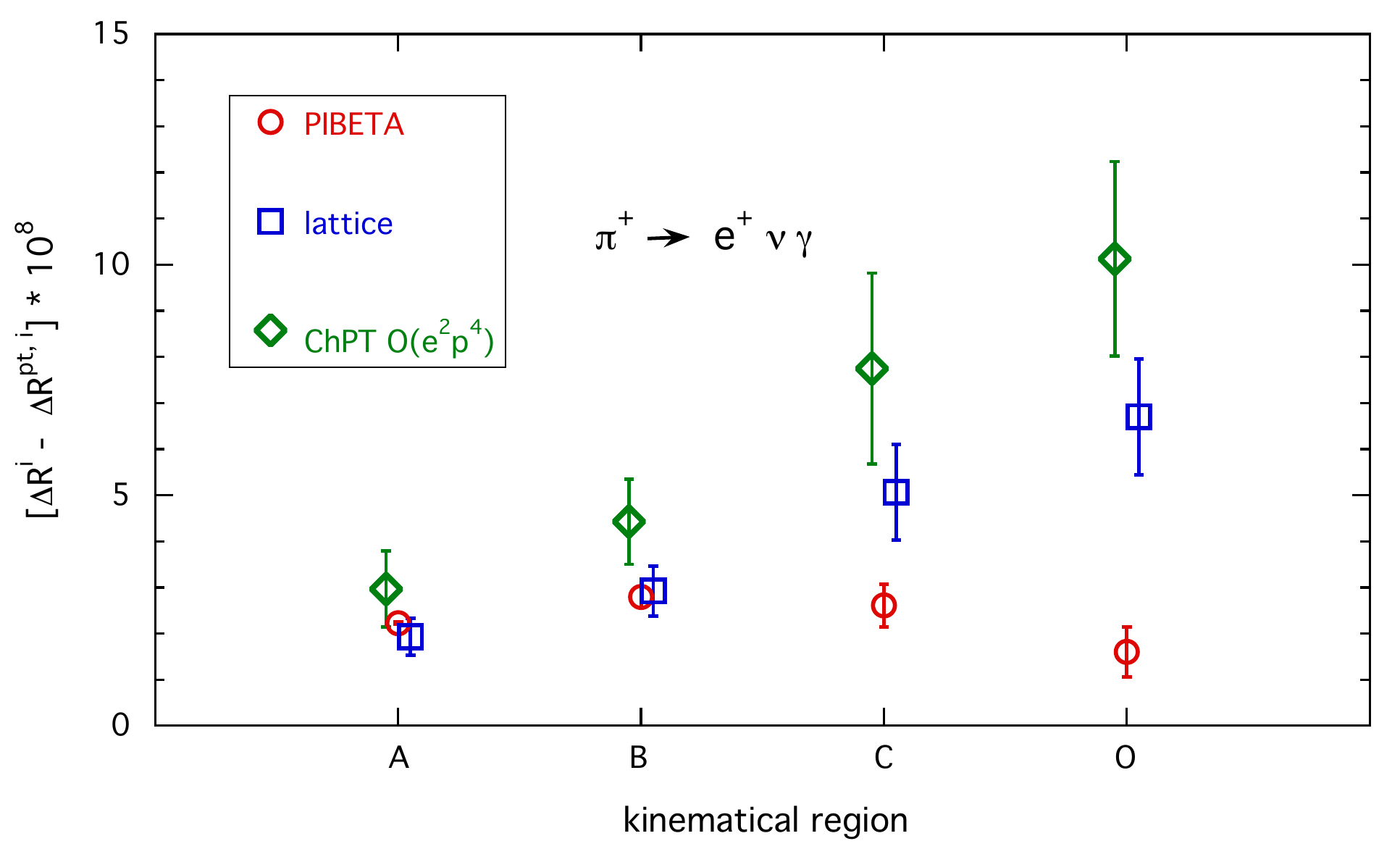}
\end{center}
\vspace{-0.75cm}
\caption{\it \small Comparison of the PIBETA experimental data\,\cite{Bychkov:2008ws} with the pt contribution subtracted,  ($\Delta R^{\mathrm{exp}, i} - \Delta R^{\mathrm{pt}, i}$) (red circles), with the theoretical predictions ($\Delta R^{\mathrm{th}, i} - \Delta R^{\mathrm{pt}, i}$) (blue squares), evaluated with the vector and axial form factors of Ref.\,\cite{Desiderio:2020oej} given in Eqs.\,(\ref{eq:FVA_linear})\,-\,(\ref{eq:FA_kaon}), for the four kinematical regions adopted in the PIBETA experiment on $\pi^+ \to e^+ \nu \gamma$ decays. The green diamonds correspond to the prediction of ChPT at order ${\cal{O}}(e^2p^4)$, based on the vector and axial form factors given in Eq.\,(\ref{eq:ChPT}).}
\label{fig:PIBETA}
\end{figure}

Using the form factors (\ref{eq:FVA_linear}) with the parameters given in Eqs.\,(\ref{eq:FV_pion}) and (\ref{eq:FA_pion}), the INT contribution $\Delta R^{\mathrm{INT}, i}$ is negligible in all the kinematical regions and the SD term $\Delta R^{\mathrm{SD}, i}$ is dominant only in region $A$, while in the other kinematical regions the pt term $\Delta R^{\mathrm{pt}, i}$ dominates.
Therefore, in order to better highlight the SD contribution we subtract from the experimental data the pt contribution, which is a purely kinematical effect and does not introduce any uncertainty. 
The values of $\Delta R^{\mathrm{pt}, i}$, of our non-perturbative predictions for $\Delta R^{\mathrm{th}, i} - \Delta R^{\mathrm{pt}, i}$ and of the {\it subtracted} experimental value $\Delta R^{\mathrm{exp}, i} - \Delta R^{\mathrm{pt}, i}$ are collected in Table\,\ref{tab:PIBETA} and shown in Fig.\,\ref{fig:PIBETA}.
In Table~\ref{tab:PIBETA} the last column shows the ChPT predictions at order ${\cal{O}}(e^2p^4)$, based on the vector and axial form factors given in Eq.\,(\ref{eq:ChPT}) with $m_\pi / f_\pi = 139.6~{\rm MeV} / 130.4~{\rm MeV}$.

\begin{table}[!hbt]
\renewcommand{\arraystretch}{1.2}	 
\begin{center}	
{\footnotesize
\begin{tabular}{||c|c|c|c||c||c|c||c|c||c||}
\hline
region & $E_\gamma$ & $E_e$ & $\theta_{e \gamma}$ & $\Delta R^{\mathrm{exp}, i}$ & $\Delta R^{\mathrm{pt}, i}$ & $(\Delta R^{\mathrm{exp}, i} -  \Delta R^{\mathrm{pt}, i})$ 
           & $\Delta R^{\mathrm{SD}, i}$ & $(\Delta R^{\mathrm{th}, i} - \Delta R^{\mathrm{pt}, i})$ & ChPT\\
\hline
A & $>50$ & $>50$    & $ >40^\circ$ & $2.614 \pm 0.021$ & $0.385$ & $2.229 \pm 0.021$ & $1.94 \pm 0.40$ & $1.93 \pm 0.40$ & $2.97 \pm 0.82$\\
B & $>50$ & $>10$    & $>40^\circ$ & $14.46 \pm 0.22$ & $11.66$ & $2.80 \pm 0.22$ & $3.01 \pm 0.54$ & $2.93 \pm 0.54$ & $4.43 \pm 0.92$\\
C & $>10$ & $>50$    & $>40^\circ$ & $37.69 \pm 0.46$ & $35.08$ & $2.61 \pm 0.46$ & $5.07 \pm 1.03$ & $5.07 \pm 1.04$ & $7.75 \pm 2.07$\\
O & $>10$ & $>m_e$ & $>40^\circ$ & $73.86 \pm 0.54$ & $72.26$ & $1.60 \pm 0.54$ & $6.87 \pm 1.26$ & $6.70 \pm 1.26$ & $10.13 \pm 2.11$\\
\hline
\end{tabular}
}
\end{center}
\renewcommand{\arraystretch}{1.0}
\caption{\it \small Values of the PIBETA experimental results $\Delta R^{\mathrm{exp}, i}$\,\cite{Bychkov:2008ws}, of the pt contribution $\Delta R^{\mathrm{pt}, i}$, of the quantity ($\Delta R^{\mathrm{exp}, i} - \Delta R^{\mathrm{pt}, i}$) and of the theoretical predictions $\Delta R^{\mathrm{SD}, i}$ and ($\Delta R^{\mathrm{th}, i} - \Delta R^{\mathrm{pt}, i}$), evaluated with the vector and axial form factors of Ref.\,\cite{Desiderio:2020oej} given in Eqs.\,(\ref{eq:FVA_linear})\,-\,(\ref{eq:FA_kaon}), corresponding to the four kinematical regions adopted in the PIBETA experiment on $\pi^+ \to e^+ \nu \gamma$ decays. Energies and branching ratios are given in units of MeV and $10^{-8}$, respectively. In the kinematical region $A$ the constraint $\theta_{e \gamma} > 40^\circ$ is automatically satisfied\,\cite{Bychkov:2008ws}. The last column shows the prediction of ChPT at order ${\cal{O}}(e^2p^4)$, based on the vector and axial form factors given in Eq.\,(\ref{eq:ChPT}).}
\label{tab:PIBETA}
\end{table} 

It can be seen that in the kinematical regions $A$ and $B$ the agreement between theory and experiment is good, while for the kinematical regions $C$ and $O$, where the ChPT predictions at order ${\cal{O}}(e^2p^4)$ also differ significantly from the measurements, a tension occurs at a level of about 2.2 and 4.1 standard deviations respectively\,\footnote{A tension of about 2.8 standard deviations is also present between our predictions and the older experimental data from ISTRA Collaboration\,\cite{Bolotov:1990yq}. There the kinematical cuts $E_\gamma > 21$\,MeV and $E_e > (70 - 0.8 E_\gamma)$\,MeV were applied, which implies that $\theta_{e \gamma} > 60^\circ$.}.
Since the kinematical regions defined by the PIBETA experiment largely overlap with each other, the precise knowledge of the covariance matrix is crucial for obtaining any reasonable estimate of the global reduced $\chi^2$-variable (see Eq.\,(\ref{eq:chi2_reduced})). 
Thus, in the absence of the experimental covariance matrix we do not provide any estimate of $\chi_{\mathrm{red}}^2$ for the PIBETA experiment.

Possible contributions in the PIBETA kinematics arising from tensor interactions beyond the SM have been discussed in the literature (see e.g. Refs.\,\cite{Chizhov:2004tu,Mateu:2007tr} and references therein).
In Ref.\,\cite{Unterdorfer:2008zz} the impact of ${\cal{O}}(e^2p^6)$ terms was estimated using also the large $N_c$ expansion within ChPT and found to be at the level of about 15\% on the axial form factor. 
Such a contribution led to a better agreement with the PIBETA data and to the conclusion that the addition of tensor interactions was not needed.
Our lattice results for the kinematical region $C$ and possibly also for the kinematical region $O$ might open again the issue of the role of possible flavor-changing interactions beyond the $V - A$ theory in radiative pion decays.

\section{SM fit to the experimental data}
\label{sec:SM_fit}

The results obtained in the previous sections naturally raise the issue of whether the vector and axial form factors can be modified in such a way as to significantly reduce the discrepancies with all the experimental data while staying within the SM. To this end the KLOE, E787, ISTRA+ and OKA data can be fitted simultaneously since they concern kaon decays, while only the PIBETA experiment measures the pion decay rates. We stress that the discussion in this section assumes the validity of the SM in general, and lepton-flavour universality in particular, allowing us to combine data from kaon decays into electrons and muons.

For radiative kaon decays we observe that:
\begin{itemize}
\item the KLOE data include values of $x_\gamma$ in the range from approximately $0.04$ to about $1.0$. At large values of $x_\gamma$ the data are mainly governed by the form factor $F^+(x_\gamma)$, while at lower values of $x_\gamma$ the data are also moderately sensitive to the form factor $F^-(x_\gamma)$;
\item the E787 data cover a range of values of $x_\gamma$ from approximately $0.36$ to about $0.96$. They are sensitive to the form factor $F^+(x_\gamma)$ at large values of $x_\gamma$ and to a lesser extent also to the form factor $F^-(x_\gamma)$ at lower values of $x_\gamma$;
\item the ISTRA+ and OKA data include values of $x_\gamma$ in the range $0.05 \lesssim x_\gamma \lesssim 0.60$ and they are sensitive to the form factor $F^-(x_\gamma)$ at large values of $x_\gamma$.
\end{itemize}

In fitting the kaon data we adopt a simple linear parameterization of the form factors $F_\pm(x_\gamma)$, suggested by our lattice results, namely 
 \be
     \label{eq:linear_fit}
     F_\pm(x_\gamma) = \widetilde{C}_\pm + \widetilde{D}_\pm x_\gamma ~ , ~
 \ee
where the four quantities $\widetilde{C}_\pm$ and $\widetilde{D}_\pm$ are now treated as free parameters.

\begin{figure}[htb!]
\begin{center}
\includegraphics[scale=0.85]{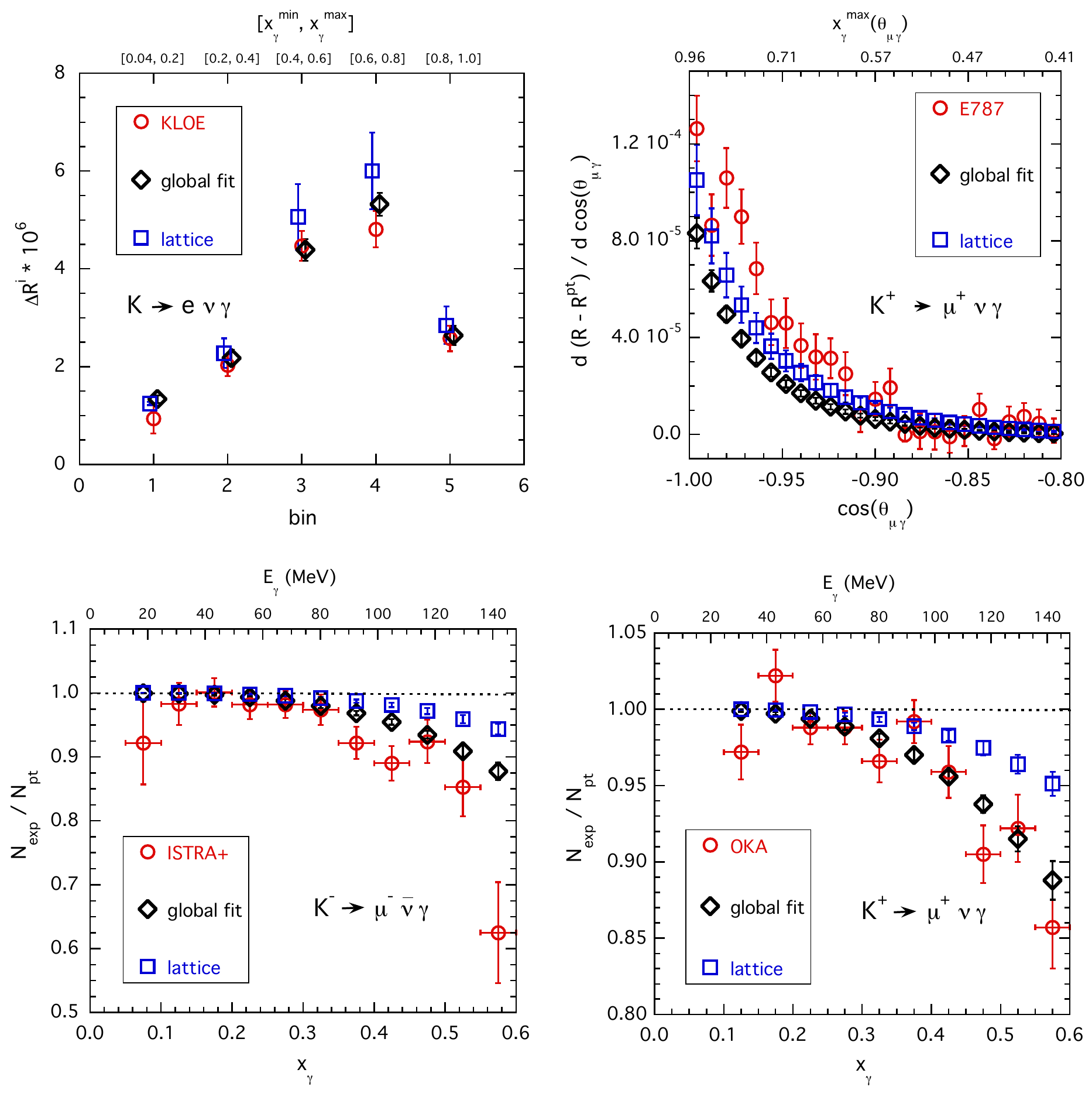}
\end{center}
\vspace{-0.85cm}
\caption{\it \small Results of the global SM fit (black diamonds) applied to the KLOE\,\cite{Ambrosino:2009aa}, E787\,\cite{Adler:2000vk}, ISTRA+\,\cite{Duk:2010bs} and OKA\,\cite{Kravtsov:2019amb} data (red circles) adopting the linear parameterization (\ref{eq:linear_fit}) for the form factors $F^+(x_\gamma)$ and $F^-(x_\gamma)$. The blue squares represent the theoretical SM predictions evaluated with the lattice form factors determined in Ref.\,\cite{Desiderio:2020oej}.}
\label{fig:SMfit}
\end{figure}

A total of 51 experimental data points (5 points from KLOE, 25 points from E787, 11 points from $\mbox{ISTRA+}$ and 10 points from OKA) are then fitted using the form factors (\ref{eq:linear_fit}) adopting a standard $\chi^2$-minimization procedure with a bootstrap sample of 5000 events generated to propagate the uncertainties of the experimental data and giving the same weight to each of the four experiments.
We remind the reader that for the various kaon experiments the correlation matrices of the data are not available.
Therefore, in our fitting procedure the experimental data are treated as uncorrelated.
The quality of the best fit is poor: the optimal value of $\chi^2 / \mbox{(no.\,of\,points)}$ is equal to $1.3, 5.3, 3.1$ and $2.2$ for the KLOE, E787, ISTRA+ and OKA data, respectively.
The comparison of the results of the global SM fit with all the experimental data is shown in Fig.\,\ref{fig:SMfit}.
The largest tension occurs for the E787 data and is a consequence of the simultaneous presence of the KLOE data, as will be explained below.

The values found for the four parameters appearing in Eq.\,(\ref{eq:linear_fit}) are determined to be
\bea
     \label{eq:F+_kaon}
     && \widetilde{C}_+ = 0.134 \pm 0.012 ~ , ~~ \widetilde{D}_+ = -0.002 \pm 0.019 ~ , ~ \\[2mm] 
     \label{eq:F-_kaon}
     && \widetilde{C}_- = 0.157 \pm 0.049 ~ , ~~  \widetilde{D}_- = -0.003 \pm 0.102 ~ , ~  \qquad
\eea
while for comparison the values of the same parameters corresponding to the lattice form factors (\ref{eq:FV_kaon}) and (\ref{eq:FA_kaon}) are
\bea
     \label{eq:F+_kaon_lattice}
     && C_+ = 0.161 \pm 0.013 ~ , ~~ D_+ = -0.025 \pm 0.011 ~ , ~ \\[2mm] 
     \label{eq:F-_kaon_lattice}
     &&C_- = 0.087 \pm 0.013 ~ , ~~  D_- = -0.023 \pm 0.014 ~ . ~  \qquad
\eea
The corresponding correlation matrices are presented in Tables\,\ref{tab:corr_exps} and\,\ref{tab:corr_lattice}.

\begin{table}[!hbt]
\renewcommand{\arraystretch}{1.2}
\begin{center}
\begin{minipage}{0.475\linewidth}	
\begin{tabular}{||c||c|c|c|c||}
\hline
 & ~~~$\widetilde{C}_+$~~~ & ~~~$\widetilde{C}_-$~~~ & ~~~$\widetilde{D}_+$~~~ & ~~~$\widetilde{D}_-$~~~ \\
\hline \hline
 $\widetilde{C}_+$ & ~1.0      & -0.393   & -0.975   & ~0.337 \\ \hline 
 $\widetilde{C}_-$  & -0.393   & ~1.0      & ~0.379  & -0.962 \\ \hline
 $\widetilde{D}_+$ & -0.975   & ~0.379  & ~1.0      & -0.331 \\ \hline
 $\widetilde{D}_-$  & ~0.337  & -0.962   & -0.331   & ~1.0 \\ \hline
\hline
\end{tabular}
\caption{\it \small Correlation matrix for the parameters $\widetilde{C}_+$, $\widetilde{C}_-$, $\widetilde{D}_+$ and $\widetilde{D}_-$ (see Eqs.\,(\ref{eq:F+_kaon}) and (\ref{eq:F-_kaon})) of the linear parameterization (\ref{eq:linear_fit}) adopted for the SM fit of the KLOE, E787, ISTRA+ and OKA data.}
\label{tab:corr_exps}
\end{minipage}
\hfill
\begin{minipage}{0.475\linewidth}	
\begin{tabular}{||c||c|c|c|c||}
\hline
 & ~~~$C_+$~~~ & ~~~$C_-$~~~ & ~~~$D_+$~~~ & ~~~$D_-$~~~ \\
\hline \hline
 $C_+$ & ~1.0     & ~0.087 & -0.703  & -0.118 \\ \hline 
 $C_-$  & ~0.087 & ~1.0     & -0.196  & -0.693 \\ \hline
 $D_+$ & -0.703  & -0.196  & ~1.0     & ~0.297 \\ \hline
 $D_-$  & -0.118  & -0.693  & ~0.297 & ~1.0 \\ \hline
\hline
\end{tabular}
\caption{\it \small Correlation matrix for the parameters $C_+$, $C_-$, $D_+$ and $D_-$ (see Eqs.\,(\ref{eq:F+_kaon_lattice}) and (\ref{eq:F-_kaon_lattice})) of the linear parameterization of the lattice form factors $F^+(x_\gamma)$ and $F^-(x_\gamma)$ determined in Ref.\,\cite{Desiderio:2020oej}.}
\label{tab:corr_lattice}
\end{minipage}
\end{center}
\renewcommand{\arraystretch}{1.0}
\end{table}
Note that the dependence on the form factor $F^+(x_\gamma)$ in the global fit to all the data is dominated by the SD$^+$ term and hence by $|F^+(x_\gamma)|$. 
We are therefore unable to determine the sign of $\widetilde{C}_+$ from the global fit alone. 
Given that both our lattice results and ChPT yield a positive value of $C_+$, we have started our minimization procedure with a positive value and subsequently always obtained positive final values of $\widetilde{C}_+$ for all the bootstrap events.

\begin{figure}[htb!]
\begin{center}
\includegraphics[scale=0.85]{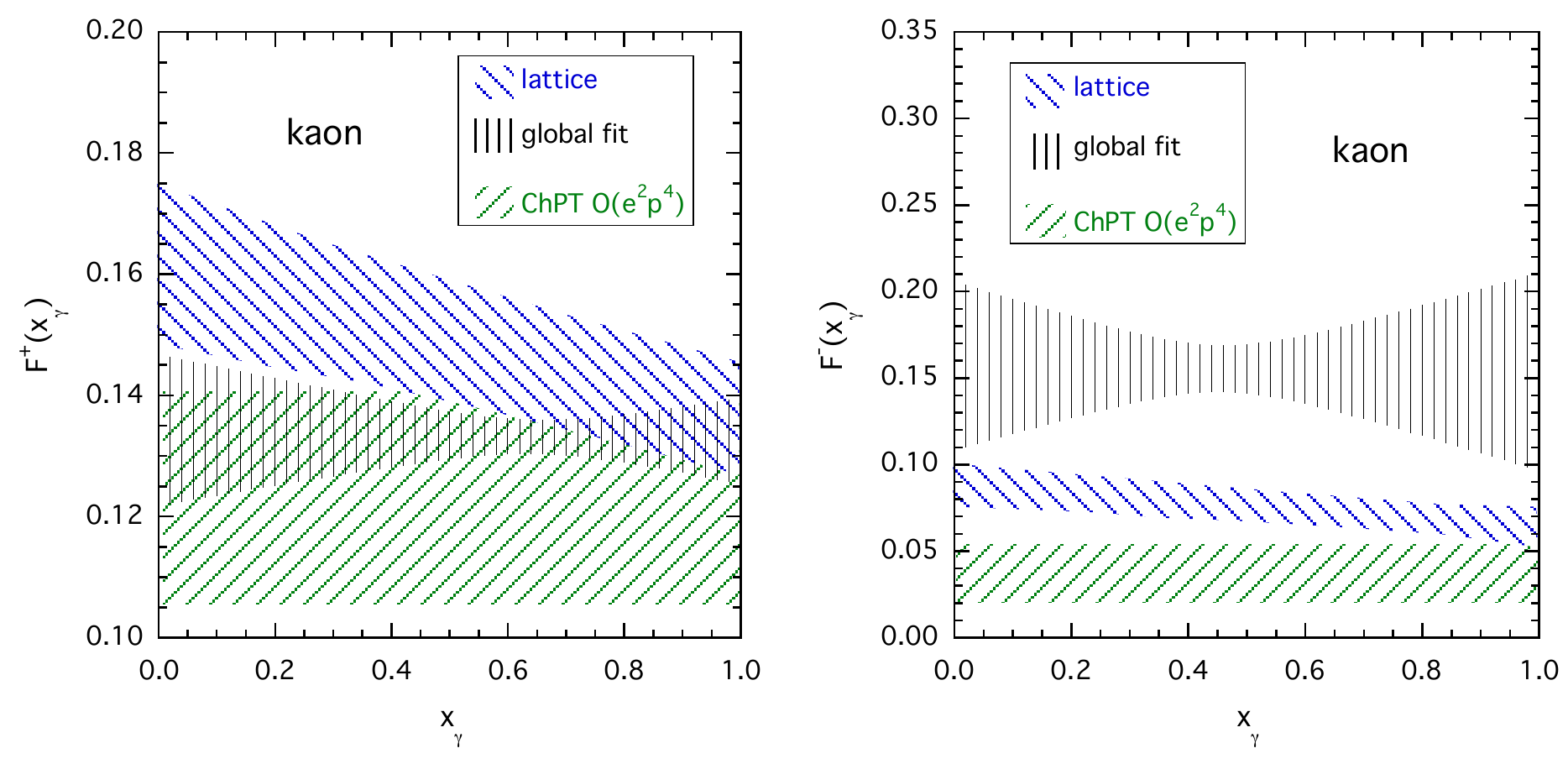}
\end{center}
\vspace{-0.85cm}
\caption{\it \small Comparison of the form factors $F^+(x_\gamma)$ (left panel) and $F^-(x_\gamma)$ (right panel), given in Eq.\,(\ref{eq:linear_fit}), obtained by the simultaneous fit of the KLOE\,\cite{Ambrosino:2009aa}, E787\,\cite{Adler:2000vk}, ISTRA+\,\cite{Duk:2010bs} and OKA\,\cite{Kravtsov:2019amb} experimental data corresponding to Eqs.\,(\ref{eq:F+_kaon})-(\ref{eq:F-_kaon}), with our lattice results from Ref.\,\cite{Desiderio:2020oej} corresponding to Eqs.\,(\ref{eq:F+_kaon_lattice})-(\ref{eq:F-_kaon_lattice}) and with the ChPT predictions at order ${\cal{O}}(e^2p^4)$ given by Eq.\,(\ref{eq:ChPT}). All the shaded areas represent uncertainties at the level of 1 standard deviation.}
\label{fig:ffs}
\end{figure}

In Fig.\,\ref{fig:ffs} the ``optimal" form factors (obtained from Eqs.\,(\ref{eq:F+_kaon}) and (\ref{eq:F-_kaon})) are compared to our lattice form factors (obtained from Eqs.\,(\ref{eq:F+_kaon_lattice}) and (\ref{eq:F-_kaon_lattice})) and to the corresponding predictions of ChPT at order ${\cal{O}}(e^2p^4)$ given by Eq.\,(\ref{eq:ChPT}).
While the discrepancy for the form factor $F^+(x_\gamma)$ is relatively mild, for $F^-(x_\gamma)$ there is a discrepancy of a factor of approximately 2 with the lattice results and even more with the ${\cal{O}}(e^2p^4)$ ChPT predictions. 
We have also explicitly checked that similar qualitative conclusions hold if different parameterizations of the $x_\gamma$ dependence of the form factors $F_\pm(x_\gamma)$ to that in Eq.\,(\ref{eq:linear_fit}) are adopted.

\begin{figure}[htb!]
\begin{center}
\includegraphics[scale=0.85]{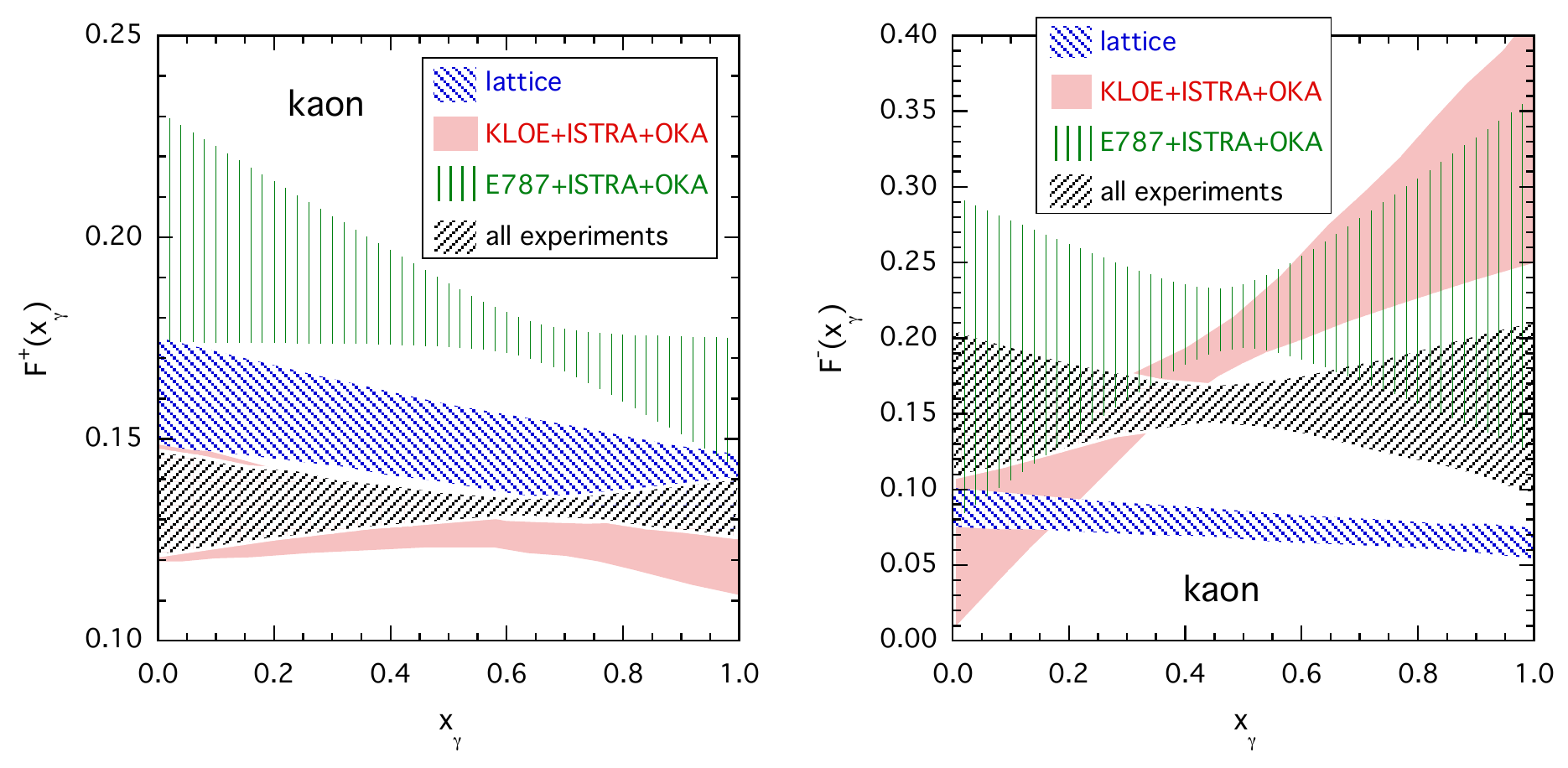}
\end{center}
\vspace{-0.85cm}
\caption{\it \small Comparison of the form factors $F^+(x_\gamma)$ (left panel) and $F^-(x_\gamma)$ (right panel), given in Eq.\,(\ref{eq:linear_fit}), obtained by the fitting either KLOE\,\cite{Ambrosino:2009aa}, ISTRA+\,\cite{Duk:2010bs} and OKA\,\cite{Kravtsov:2019amb} data (red shaded areas) or E787\,\cite{Adler:2000vk}, ISTRA+\,\cite{Duk:2010bs} and OKA\,\cite{Kravtsov:2019amb} data (green shaded areas). The black shaded areas correspond to the simultaneous fit of all the experimental data from KLOE\,\cite{Ambrosino:2009aa}, E787\,\cite{Adler:2000vk}, ISTRA+\,\cite{Duk:2010bs} and OKA\,\cite{Kravtsov:2019amb}. The blue shaded areas represent our lattice results from Ref.\,\cite{Desiderio:2020oej}. All the shaded areas represent uncertainties at the level of 1 standard deviation.}
\label{fig:ffs_comp}
\end{figure}

The difficulty in performing a global fit within the SM is partly due to the inconsistent results in the form factor $F^+(x_\gamma)$ from the KLOE and E787 experiments, as discussed in Sec.\,\ref{sec:E787}. 
This is further illustrated in Fig.\,\ref{fig:ffs_comp} where the results for the form factors from the best fits are plotted omitting either the E787 data or the KLOE data and compared to the lattice results. 
The best separate fits to the KLOE and E787 data result in significantly different values of the form factor $F^+(x_\gamma)$.
It can also be seen that the optimal form factor $F^-(x_\gamma)$ always deviates significantly from our lattice results and its slope is also sensitive to the inclusion of either the KLOE or the E787 data or both.
At low values of $x_\gamma$ the KLOE data prefer smaller values of the form factor $F^-(x_\gamma)$, while the E787 data are compatible with larger ones.
This is again related to the different values of the form factor $F^+(x_\gamma)$ from the KLOE and E787 experiments shown in the left panel of Fig.\,\ref{fig:ffs_comp}.
At large values of $x_\gamma$ the form factor $F^-(x_\gamma)$ is mainly governed by the ISTRA+ and OKA data\footnote{The inclusion of the E787 data has two main consequences on the optimal form factors corresponding to the KLOE+ISTRA+OKA analysis: 1) at large $x_\gamma$ the form factor $F^+(x_\gamma)$ increases and correspondingly the form factor $F^-(x_\gamma)$ should decrease to keep unchanged the sum SD$^+$ + SD$^-$ governed by the KLOE data; 2) at low $x_\gamma$ the E787 data for $\mbox{cos}(\theta_{\mu \gamma}) \gtrsim -0.9$ (see Fig.\,\ref{fig:E787}) require larger values of $F^-(x_\gamma)$ to compensate the SD$^+$ + INT$^+$ contribution. The above features produce the flattening of $F^-(x_\gamma)$ observed in Fig.\,\ref{fig:ffs_comp} when all the experiments are considered.}.
Finally note that the extraction of the form factor $F^+(x_\gamma)$ from the KLOE data is affected by the simultaneous inclusion of the ISTRA+ and OKA data at low values of $x_\gamma$ (compare the red striped area in the right panel of Fig.\,\ref{fig:KLOE} with the red shaded area in the left panel of Fig.\,\ref{fig:ffs_comp}).

\section{Conclusions}
\label{sec:conclusions}

We have presented a comparison of our theoretical predictions with the existing experimental data on the radiative leptonic decays $K\to e \nu_e \gamma$ from the KLOE collaboration\,\cite{Ambrosino:2009aa}, $K\to \mu\nu_\mu\gamma$ from the E787, ISTRA+ and OKA collaborations\,\cite{Adler:2000vk,Duk:2010bs,Kravtsov:2019amb} and $\pi\to e\nu_e\gamma$ from the PIBETA experiment\,\cite{Bychkov:2008ws}.
The theoretical predictions are based on our recent non-perturbative determinations of the vector and axial-vector form factors corresponding to the emission of a real photon, using lattice QCD+QED simulations at leading order in the electromagnetic coupling, $O(\alpha_{\mathrm{em}})$,  in the electroquenched approximation\,\cite{Desiderio:2020oej}. 

We find good consistency between our theoretical predictions and the experimental results from the KLOE experiment on  $K\to e\nu\gamma$ decays\,\cite{Ambrosino:2009aa}, but a discrepancy at the level of about 2 standard deviations for the data at large $x_\gamma$ from the E787 experiment on $K\to \mu\nu\gamma$ decays. Indeed the results from the two experiments do not agree. 
We also find differences of up to 3\,-\,4 standard deviations at large photon energies in the comparison of our predictions with the E787, ISTRA+ and OKA data on radiative kaon decays as well as for some kinematical regions of the PIBETA experiment on the radiative pion decay.

We have also performed a simultaneous fit of the KLOE, E787, ISTRA+ and OKA experimental data on the radiative kaon decays staying within the SM and adopting the linear ansatz in Eq.\,(\ref{eq:linear_fit}) for the SD form factors $F_\pm(x_\gamma)$, as suggested by the lattice results of Ref.\,\cite{Desiderio:2020oej}. 
The quality of the fit is poor because, as mentioned above, the KLOE and E787 data cannot be reproduced simultaneously in terms of the same form factor $F^+(x_\gamma)$. 
We find a particularly significant discrepancy between our predictions and the experimental data for the form factor $F^-(x_\gamma)$.

These conclusions call for improvements in the determination of the structure-dependent form factors $F^+(x_\gamma)$ and $F^-(x_\gamma)$ from both experiment and theory. 
In this respect, we look forward to the results from the analysis of the NA62 experiment on the $K_{e2\gamma}$ decay, which is in progress and which is expected to provide the most precise determination of $|F^+(x_\gamma)|$~\cite{NA62}. If the results from NA62 confirm that there is a discrepancy between the form factors obtained from decays into electrons and those obtained from decays into muons from the E787 experiment, this would provide a motivation for better determinations also of the form factors from $K\to\mu\nu_\mu\gamma$ decays.
On the theoretical side is should be noted that the values of $F_{V,A}$ in Ref.\,\cite{Desiderio:2020oej} are the first lattice results of these quantities, so it can be expected that the precision will be improved in the next generation of computations. 

We end by repeating that it is also conceivable that the tensions observed above between the experimental data and our lattice predictions are due to the presence of new physics, such as flavor changing interactions beyond the $V-A$ couplings of the Standard Model and/or non-universal corrections to the lepton couplings. 
This possibility deserves further theoretical investigations.

\section*{Acknowledgments}

The authors warmly thanks D.\,Giusti for a careful check of the implementation of the kinematical cuts. We gratefully acknowledge discussions with members of the experimental collaborations about their data and results and we thank in particular B.\,Sciascia and T.\,Spadaro from KLOE\,\cite{Ambrosino:2009aa}, M.R.\,Convery from E787\,\cite{Adler:2000vk} and
M.\,Bychkov from PIBETA\,\cite{Bychkov:2008ws}. We also thank V.\,Kravtsov, V.\,Duk and V.\,Obraztsov for supplying us with the kinematical cuts of the OKA experiment\,\cite{Kravtsov:2019amb} and A.\,Romano for discussions about the status of the ongoing analysis by the NA62 experiment.
We acknowledge PRACE for awarding us access to Marconi at CINECA, Italy under the grant Pra17-4394 and CINECA for the provision of CPU time under the specific initiative INFN-LQCD123.
C.T.S.~was partially supported by an Emeritus Fellowship from the Leverhulme Trust and by STFC (UK) grants\,ST/P000711/1 and ST/T000775/1.
N.T.~and R.F.~acknowledge the University of Rome Tor Vergata for the support granted to the project PLNUGAMMA.
F.S.~and S.S.~are supported by the Italian Ministry of University and Research (MIUR) under grant PRIN 20172LNEEZ. 
F.S.~is supported by INFN under GRANT73/CALAT.

\end{document}